%% file: CMD_arXiv_v3.tex
\documentclass[11pt,a4paper]{article}

\usepackage[english]{babel}
\usepackage{titling}\droptitle1cm
\usepackage{fancyhdr}

\usepackage[authoryear]{natbib}

\usepackage[unicode=true,pdfusetitle, bookmarks=true, breaklinks=true,pdfborder={0 0 1},backref=false,colorlinks=true]{hyperref}
\addto\extrasenglish{}
\addto\extrasenglish{}
\addto\extrasenglish{}

\newcommand{\samecolor}{0,0,.8}
\hypersetup{linkcolor=[rgb]{\samecolor},citecolor=[rgb]{\samecolor},urlcolor=[rgb]{\samecolor},bookmarksnumbered=true,pdfpagelayout=SinglePage}

\newcommand{\appref}[1]{\hyperref[#1]{Appendix \ref*{#1}}}

\usepackage[]{caption} 

\usepackage{amsmath,amssymb,amsthm,url,graphicx,
	booktabs,multirow,fullpage}
\usepackage[dvipsnames]{xcolor}
\usepackage[doublespacing]{setspace}
\usepackage{verbatim}
\title{Score-based calibration testing for multivariate forecast distributions\thanks{We thank Timo Dimitriadis, Tobias Fissler, Ana Galv\~ao, Tilmann Gneiting, Johanna Ziegel and seminar participants at Deutsche Bundesbank, Heidelberg Institute of Theoretical Studies, Karlsruhe Institute of Technology, University of Cologne and participants of the International Symposium on Forecasting 2021 (virtual), the Statistische Woche 2021 (virtual), the Bundesbank/EABCN conference ``Challenges in Empirical Macroeconomics since 2020'' in Eltville, the DAGStat Conference 2022 in Hamburg, the International Association for Applied Econometrics 2022 Annual Conference in London, and the 2022 meeting of the Committee for Econometrics of the German Economic Association for helpful comments. We acknowledge support by the state of Baden-W\"urttemberg through bwHPC. Marc-Oliver Pohle is grateful for support by the Klaus Tschira Foundation, Germany. The views expressed in this paper are those of the authors and do not necessarily coincide with the views of the Deutsche Bundesbank or the Eurosystem.}}

\newcommand{\E}{\mathbb{E}}

\newtheorem{Definition}{Definition}[section]

\author{Malte Kn\"uppel\thanks{Deutsche Bundesbank, e-mail: malte.knueppel@bundesbank.de} \and Fabian Kr\"uger\thanks{Karlsruhe Institute of Technology, e-mail: fabian.krueger@kit.edu} \and Marc-Oliver Pohle\thanks{Heidelberg Institute for Theoretical Studies, e-mail: marc-oliver.pohle@h-its.org}}
\date{\today}


\begin{document}

\maketitle
\thispagestyle{empty}


\begin{abstract}
	Calibration tests based on the probability integral transform (PIT) are routinely used to assess the quality of univariate distributional forecasts. However, PIT-based calibration tests for multivariate distributional forecasts face various challenges. We propose two new types of tests based on proper scoring rules, which overcome these challenges. They arise from a general framework for calibration testing in the multivariate case, introduced in this work. The new tests have good size and power properties in simulations and solve various problems of existing tests. We apply the tests to forecast distributions for macroeconomic and financial time series data.


\end{abstract}

\newpage
\onehalfspacing
\setcounter{page}{1}
\section{Introduction}

Decision-making and planning for the future requires good forecasts, often not only of the mean of a variable of interest, but also of other features of its probability distribution, for example the variance or certain quantiles. Ideally, probabilistic forecasts, that is forecasts of the full probability distribution, are provided. Consequently, there has been a gradual shift from point to probabilistic forecasting throughout scientific disciplines and fields of application  (\citealp{Gneiting2008}; \citealp{GneitingKatzfuss2014}). At the same time, decision-making often requires forecasts of multiple variables of interest. Examples include forecasts of macroeconomic variables like GDP growth, inflation and an interest rate in economics, returns of a possibly large group of assets in finance or precipitation amounts at different locations in meteorology. Probabilistic forecasts then take the form of multivariate density or distribution functions. Such forecasts are, for instance, studied by  \cite{Koop2013}, \cite{Chan2020}, \cite{McAlinnEtAl2020} and \cite{CarrieroEtAl2021}
in macroeconomics, by \cite{Diksetal14}, \cite{Cataniaetal2018}  and \cite{Guptaetal2020} in finance, and by \cite{ClarkEtAl2004} and \cite{HeinrichEtAl2021} in meteorology.

When evaluating multivariate forecasts, it is crucial to use methods that evaluate the forecasts for all variables jointly. Evaluating each variable of interest separately would ignore the dependencies between them, which are usually of prime importance and the reason for multivariate forecasting in the first place. Forecast evaluation methods can be divided into methods for relative and absolute evaluation. Relative forecast evaluation is concerned with comparing the accuracy of different forecasts, while absolute forecast evaluation checks if a forecast fulfills certain desirable criteria. The comparison of multivariate probabilistic forecasts in terms of their accuracy, measured by the expected score, carries over straightforwardly from the univariate case. Popular proper scoring rules for multivariate forecast distributions are, for example, the log score and the energy score \citep{GneitingRaftery2007}. Absolute forecast evaluation usually amounts to examining calibration, that is the `statistical consistency between the distributional forecasts and the observations' \citep{Gneiting2007Sharpness}. For univariate probabilistic forecasts, calibration is commonly assessed by checking uniformity of probability integral transforms \citep[PITs, see][]{Dawid1984,Diebold-et-al-98,Gneiting2007Sharpness}. The PIT evaluates the forecaster's cumulative distribution function (CDF) at the corresponding observation.\footnote{For a formal expression of the PIT, see Definition \ref{def:PIT} below.}
Two separate strands of literature (in econometrics on the one hand and meteorology and statistics on the other) try to generalize this approach from the univariate setting, but interestingly in quite different ways.

In econometrics, \citet{Diebold-et-al-98} and \citet{Diebold-et-al-99b} popularized an approach alluded to by \citet{Smith-85}, which uses the decomposition of the $d$-dimensional forecast distribution into $d$ conditional distributions and the calculation of the PITs for each of these $d$ (univariate) conditional distributions. The null hypothesis of optimality or ideal calibration implies that each PIT follows an independent $U(0,1)$ distribution, and this insight can be used to construct tests. Yet, such tests suffer from the drawbacks that there is no natural ordering of the variables in the decomposition, and that the decomposition essentially requires the multivariate forecast distribution to be available in closed form. While \citeauthor{DovernManner2020}'s \citeyearpar{DovernManner2020} tests address the former problem, the latter problem has remained unresolved in econometrics.

The meteorological literature traditionally focuses on graphical diagnostic checks rather than on statistical testing. The usual diagnostic check for calibration of multivariate distributional forecasts proposed by \cite{GneitingEtAl2008} consists of two steps: First, a so-called prerank function is applied to reduce the dimension from $d$ to 1. Second, a univariate PIT or a rank histogram is constructed from these univariate quantities and checked visually for uniformity. Two popular prerank functions aim for a generalization of the univariate PIT: The multivariate rank histogram or Copula-PIT (\citealp{GneitingEtAl2008}; \citealp{ZiegelGneiting2014}) and the average rank histogram (\citealp{ThorarinsdottirEtAl2016}). \cite{wilks2017} and \cite{thorarinsdottir2018} provide overviews and comparisons of different prerank functions, while \cite{ziegel2015} discusses them from a theoretical perspective. \cite{GneitingEtAl2008} also propose an alternative approach: They suggest to check standard uniformity of the Box density ordinate transform by \cite{Box-80}.

In our work, we consider the null hypothesis of auto-calibration as introduced by \cite{Tsyplakov2011} and \cite{GneitingRanjan2013} in the case of univariate probabilistic forecasts. Auto-calibration denotes optimality of the forecasts with respect to the information contained in the forecasts themselves. Intuitively, a forecast distribution is auto-calibrated if it cannot be improved by any form of transformation, for instance, by shifting its location or increasing its spread. The notion of auto-calibration generalizes readily to the multivariate case. This is in contrast to the weaker null hypothesis of probabilistic calibration (\citealp{Gneiting2007Sharpness}), or equivalently, PIT uniformity, that is commonly employed in the univariate setting.

It is clear that information needs to be condensed in order to move from the null hypothesis involving multivariate forecast distributions and conditional distributions of outcomes to a one-dimensional test statistic. Using a new general framework for testing auto-calibration of multivariate probabilistic forecasts, we clarify the choices involved in constructing such tests. By uncovering the implicit choices of existing tests, their relations become easier to understand. Moreover, this framework facilitates the search for appropriate choices in different situations and hence allows the construction of new tests.

More specifically, our framework illustrates that the construction of a test involves two steps: First, a dimensionality reduction function $g(F,Y)$, a generalization of the prerank function from the meteorological literature, needs to be chosen. This function takes the outcome $Y$ and the forecast distribution $F$ as inputs and maps them to the real line, i.e.\ it reduces the dimensionality of the problem from $d$ to 1. 
The second step then consists in the choice of a testing principle, with a PIT uniformity test being a natural choice: One tests whether the forecast distribution for the univariate quantity $g(F,Y)$ implied by $F$ yields uniformly distributed PIT values. Many of the tests used in the literature (e.g., the tests by \citealp{GneitingEtAl2008}, \citealp{ThorarinsdottirEtAl2016}, \citealp{ZiegelGneiting2014} and \citealp{DovernManner2020}) fall under this PIT-based approach and merely differ in their choice of dimensionality reduction function. An alternative testing principle for the second step compares the realized value of the dimensionality reduction function, $g(F,Y)$, with the expected value under the forecast distribution, $\E_F[g(F,X)]$, where $X \sim F$, in terms of their means. This only requires a simple $t$-test.\footnote{The regression-based test of \cite{WeiEtAl2017} is based on the same idea, but requires restrictive assumptions, namely independent an identically distributed normal random vectors.} Temporal dependence can be addressed by using heteroscedasticity-and-autocorrelation-consistent (HAC) approaches for both testing principles considered.

The choice of the dimensionality reduction function $g$ is crucial for the performance of the tests. We propose to use proper scoring rules as dimensionality reduction functions because they are designed to summarize the quality of a forecast distribution and are thus sensitive to all of its facets. The resulting score-based calibration tests are powerful, straightforward to implement, interpretable and applicable in high-dimensional settings as we demonstrate in simulation studies and two case studies from macroeconomics (considering Bayesian vector autoregressive models) and finance (considering quantile regression and nonparametric copula methods). Further, using the energy score as dimensionality reduction function, they are applicable in settings where the forecast distribution is not available in closed form, but is represented by a sample, for example obtained by Markov chain Monte Carlo (MCMC) methods. Our tests are also nicely interpretable as they essentially compare the uncertainty predicted by the forecasting model to the realized uncertainty and thus uncover over- or underconfident forecasts. This also allows us to introduce diagnostic plots, which may give further insights into forecast performance and directions of improvement. The plots naturally complement the tests, but may also serve as stand-alone graphical tools. Our recommended test implementation, summarized for convenience in the following table, depends on whether power or interpretability is of higher priority to the user, and on whether the forecast distribution is available in closed form. When used in conjunction with scoring rules as dimensionality reduction functions we call the PIT-based approach \emph{generalized Box transform (GBT) test} and the approach employing expected values \emph{entropy test} (for reasons detailed in Section \ref{sec:score-based_tests}):\\

\begin{table}[!htbp]
	\centering
	\begin{tabular}{llll}
		Focus on & Closed-form forecast & Recommended test version & Implementation\\
		&  density available? & & reference  \\ \toprule
		Power & Yes & GBT test, log score & $\hat U_{\text{LS}, t}$,~ eq. (\ref{eqn:pit_ls})\\
		Power & No & GBT test, energy score & $\hat U_{\text{ES}, t}$,~ eq. (\ref{eqn:pit_es})  \\
		Interpretability & Yes & Entropy test, log score & $\hat D_{\text{LS}, t}$,~ eq. (\ref{eqn:dhat_logscore})\\
		Interpretability & No & Entropy test, energy score & $\hat D_{\text{ES}, t}$,~ eq. (\ref{eqn:dhat_es})\\ \bottomrule
	\end{tabular}\\
\end{table}

Section \ref{sec:general_framework} introduces our general framework for calibration testing for multivariate forecast distributions, while Section \ref{sec:score-based_tests} describes the score-based tests. Section \ref{sec:simulations} presents simulation results, Section \ref{sec:applications} and Section \ref{sec:application_finance} contain case studies on macroeconomic and financial forecasting and Section \ref{sec:conclusion} concludes. R-code to implement our proposed methods is available at \url{https://github.com/FK83/forecastcalibration}.

\section{A general framework for testing calibration of multivariate forecast distributions} \label{sec:general_framework}

\subsection{Auto-calibration and probabilistic calibration}

Throughout the paper we denote by $F_{V}$ the CDF of a random vector $V$, by $F_{V|\mathcal{H}}$ the (conditional) CDF of a random vector $V$ conditional on the $\sigma$-algebra $\mathcal{H}$. By $\sigma(W)$ we denote the $\sigma$-algebra generated by a random variable $W$ and when conditioning on $\sigma(W)$, we write $\bullet|W$ for $\bullet|\sigma(W)$. We abbreviate almost surely by $a.s.$.

Consider a $d$-dimensional random vector of interest $Y$ and the forecaster's information set $\mathcal{I}$, which is generated by a possibly very large vector of variables. A probabilistic forecaster's goal is to predict the conditional distribution of $Y$ given $\mathcal{I}$, $F_{Y|\mathcal{I}}$. We simply write $F$ for the forecast distribution itself.\footnote{Usually we are concerned with a time series forecasting problem where the forecaster stands at time $t-h$ and wants to forecast $h$ periods into the future and $\mathcal{I}$ represents all the information available at time $t-h$. Nevertheless, we do not use time indices in the theoretical part of this paper to avoid clutter and because the theory and tests presented here apply more generally to other classes of prediction problems as well, e.g.\ cross-sectional prediction problems. When we consider an evaluation sample, we will use the index $t$ to refer to the forecast and observation for time $t$ (or cross-sectional unit $t$), but still suppress the forecast horizon $h$.} For the purpose of forecast evaluation, a forecast-observation sample $\{(F_t,Y_t)\}_{t=1}^T$ is available.

We consider auto-calibration (\citealp{Tsyplakov2011}; \citealp{GneitingRanjan2013}) as the null hypothesis of our tests.
\begin{Definition} \label{dfn:ac}
The forecast distribution $F$ is \textbf{auto-calibrated} if it holds that
$$F_{Y|F}=F \text{ a.s.}.$$
\end{Definition}
A forecast is auto-calibrated if it is optimal relative to the information contained in itself, $\sigma(F)$. Intuitively, a user who receives an auto-calibrated forecast $F$ should use this forecast `as is', without any form of post-processing. Auto-calibration thus seems closely in line with the definition of calibration as the `statistical consistency between the distributional forecasts and the observations' \citep{Gneiting2007Sharpness}. The broader idea behind auto-calibration dates back to \cite{Theil1961} and \cite{Mincer1969} who
consider a scatter plot of realized values $Y$ (vertical axis) against mean forecasts $\hat Y$ (horizontal axis), arguing that
the observations in this diagram should scatter unsystematically around the 45 degree line. This requirement corresponds to $\mathbb{E}[Y|\hat Y] = \hat Y$, the analogue of Definition \ref{dfn:ac} for the mean functional, which can be viewed as the null hypothesis behind the widely-used Mincer-Zarnowitz (1969)\nocite{Mincer1969} regression. \cite{Tsyplakov2011} and \cite{GneitingRanjan2013} introduce auto-calibration in the context of distributional forecasts and coin the term. Importantly, auto-calibration transfers to all types of forecast distributions $F$, whether univariate or multivariate, continuous or discrete.\footnote{Furthermore, auto-calibration also transfers naturally to functionals such as quantiles and expectiles. \cite{Guler2017} extend \cite{Mincer1969} type regressions along these lines.}

Probabilistic calibration introduced by \cite{Gneiting2007Sharpness} reflects the current practice in the evaluation of univariate forecast distributions, in that checking PIT uniformity is the most common approach to absolute forecast evaluation throughout the disciplines. \cite{Rosenblatt-52} presented the following definition and uniformity result of the PIT.
\begin{Definition} \label{def:PIT}
	Let $H$ be a CDF and $W$ a univariate random variable. Then
	$$PIT_{H,W} = H(W)$$
	is the \textbf{probability integral transform (PIT)} of $H$ and $W$.\footnote{The PIT can be adapted to non-continuous distributions by introducing randomization at the discontinuities (see e.g.\ \cite{ruschendorf2009}). We focus on the continuous case here, but note that all the results in this paper continue to hold in the general case. Note that in the general statistical literature, this generalization of the PIT to arbitrary distributions is called distributional transform (see e.g.\ \citealp{ruschendorf2009}), but in the forecast evaluation literature it is usually still called PIT (see e.g.\ \citealp[Definition 2]{GneitingKatzfuss2014}).}
\end{Definition}

If $W \sim H$, the PIT uniformity result given by
$$PIT_{H,W} \sim U(0,1)$$
holds. Note that the PIT readily generalizes to the multivariate case, but the uniformity result does not (see \citealp{genest2001}), i.e.\ the distribution of a random vector plugged into its own CDF is in general unknown.

Returning to the forecasting framework introduced above, in the univariate case the PIT uniformity result implies that under auto-calibration we have $$PIT_{F,Y} | F \sim U(0,1).$$
This conditional uniformity of the PIT implies unconditional uniformity, which has been called probabilistic calibration by \citet[Definition 1(a)]{Gneiting2007Sharpness} and \citet[Definition 2.6(b)]{GneitingRanjan2013}.

\begin{Definition}
The forecast distribution $F$ is \textbf{probabilistically calibrated} if it holds that
$$PIT_{F,Y} \sim U(0,1).$$\label{dfn}
\end{Definition}

Due to the lack of a PIT uniformity result in the multivariate case, probabilistic calibration does not generalize, but nevertheless most approaches in the literature on calibration testing for multivariate forecast distributions strive for a suitable extension of this concept. We take a different route by starting from the null hypothesis of auto-calibration and deriving testable implications. We thus rather view PIT uniformity as an implication of auto-calibration, as opposed to being a notion of calibration in its own right. This subtle shift of perspective suggests novel ways of constructing multivariate tests.

\subsection{Testable implications of auto-calibration}

The defining equation of auto-calibration from Definition \ref{dfn:ac} is hard to test directly as this would involve the estimation of a multivariate conditional distribution function conditional on a large information set. Thus, we are looking for testable implications, which requires to condense information. We get rid of the first difficulty by reducing the dimensionality from $d$ to 1 by means of a dimensionality reduction function $g$ which maps the forecast-observation pair $(F,Y)$ to the real line. This step allows us to apply univariate evaluation methods to assess the forecast distribution of the univariate quantity $g(F, Y)$. The original forecast distribution $F$ implies a forecast distribution for this auxiliary forecasting problem, which we call $G$. Let $X$ be a random draw from $F$, i.e. $X \sim F$ (or equivalently $X=F^{-1}(U) \text{ with } U \sim U(0,1)$). Then $G = F_{g(F,X)|F}$, that is $G$ is a CDF-valued random variable fully determined by $F$. In the rest of the paper, often probabilities and expectations from $G$ show up, where we are dealing with a specific $g$ or $X$ and want to keep $g$ and $X$ explicit. We then write compactly $\mathbb{P}_F (g(F,X) \leq v)$ instead of $\mathbb{P} (g(F,X) \leq v|F)$ for $G(v)$ and $\E_F [g(F,X)]$ for its expectation instead of $\E [g(F,X)|F]$.
	
Auto-calibration implies that
\begin{equation} \label{eqn:AC_implication}
F_{g(F,Y)|F}=G \ a.s..
\end{equation}

As the estimation of $F_{g(F,Y)|F}$ is still practically infeasible due to the large information set $\sigma(F)$, we next get rid of the conditioning. There are two natural ways to achieve this. For the first, we use the PIT uniformity result and the law of total probability: Equation \eqref{eqn:AC_implication} can be written as
$$g(F,Y)|F \sim G \ a.s. ,$$
which, by the PIT uniformity result, is equivalent to
$$PIT_{G,g(F,Y)}|F \sim U(0,1) \ a.s..$$
By the law of total probability, this implies that
\begin{equation} \label{eqn:PIT_G}
U := PIT_{G,g(F,Y)} \sim U(0,1) \ a.s..
\end{equation}
Conveniently, testing Equation \eqref{eqn:PIT_G} amounts to testing probabilistic calibration of the implied forecast distribution $G$. To this end, common tests of standard uniformity under temporal dependence can be applied.

Another implication of auto-calibration arises when we concentrate on the means of the distributions showing up in \eqref{eqn:AC_implication}, which yields
\begin{equation} \label{eqn:entropy_regression}
\mathbb{E} [g(F,Y)|F] = \mathbb{E}_F [ g(F,X) ] \  a.s..
\end{equation}
\cite{Tsyplakov2014} points out this implication and suggests to use it for testing in the case of univariate probabilistic forecasts. We now use the law of iterated expectations to get rid of the conditioning:
\begin{equation*}
\mathbb{E} [ g(F,Y) ] = \mathbb{E} [ \mathbb{E}_F [ g(F,X) ]].
\end{equation*}
This equality represents an unbiasedness condition on the implied forecast distribution $G$. Rewriting it as
\begin{equation} \label{eqn:entropy}
\mathbb{E} [ D ] = 0 \text{ with } D = g(F,Y) - \mathbb{E}_F [ g(F,X) ],
\end{equation}
it is clear that it can be tested by a simple $t$-test using HAC standard errors.\footnote{Alternatively, one could use tests directly arising from equation \eqref{eqn:entropy_regression}, i.e.\ run a linear regression of $g(F,Y)$ on a constant and $\mathbb{E}_F [ g(F,X) ]$ and test for the constant being equal to zero and the slope coefficient being equal to one. This would be a direct generalization of the approach of \cite{WeiEtAl2017} that does not rely on a specific distributional assumption or a specific score. This approach can be more powerful in certain situations as two parameters are estimated contrary to only one in a $t$-test arising from \eqref{eqn:entropy}, but has the disadvantage that it leads to perfect collinearity if the entropy of the forecast distribution does not change and to multicollinearity if it hardly varies.
}

We call the two general approaches to testing auto-calibration of multivariate probabilistic forecasts coming out of equations \eqref{eqn:PIT_G} and \eqref{eqn:entropy} the \textit{PIT-based approach} and the $t$\textit{-test approach}. Most existing tests are special cases of these two approaches as we will discuss below.

\subsection{Example}

\begin{figure}
	\begin{tabular}{cc}
		Time plot of realized minus expected score & Histogram of score PIT  \\ 
		\includegraphics[width=.45\textwidth]{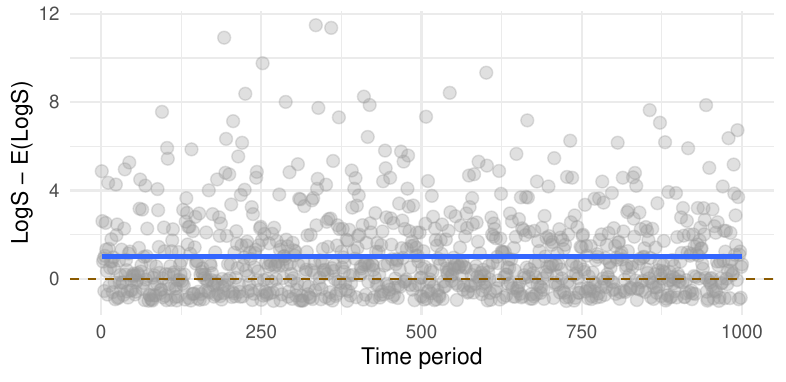} &
		\includegraphics[width=.45\textwidth]{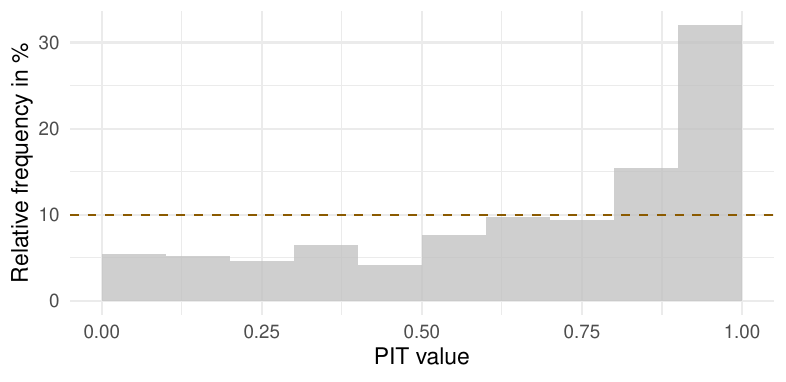} \\	
	\end{tabular}
	\caption{Illustration of the two test variants. Left: Time series plot of differences $\hat{D}_t$ between realized and expected log scores. Average difference indicated by blue line. Under auto-calibration, expected difference equals zero. Right: Histogram of PITs $\hat{U}_t$ of realized log scores. Each of the 10 bars has an expected height of 10 under auto-calibration.\label{fig:ex}}
\end{figure}

To illustrate the use of the quantities $U$ and $D$ from \eqref{eqn:PIT_G} and \eqref{eqn:entropy} for checking calibration, we consider a bivariate normal example with $Y \stackrel{\text{IID}}{\sim} \mathcal{N}\left( \mathbf{0}, \Sigma \right)$ and constant forecast distribution $F = \mathcal{N}\left( \mathbf{0}, 0.5\times \Sigma \right),$ i.e$.$ the actual variance is twice as large as the forecast variance. We further set $\Sigma = \begin{pmatrix} 1 & 0.5\\ 0.5 & 1 \end{pmatrix}$. We use the log score as a dimensionality reduction function, $g(F,Y)= -\log (f(Y))$, where $f=F^{\prime}$ and smaller scores correspond to more accurate forecasts. We obtain the distribution of $g$ by simulating from the forecast distribution $F$. Details on these issues are discussed later. We draw a sample $\{Y_t\}_{t=1}^{1000}$. The left panel of Figure \ref{fig:ex} shows a time series plot of (a simulation based estimate of) $D_t$ as defined in \eqref{eqn:entropy}.\footnote{Equation (\ref{eqn:dhat_logscore}) describes the finite sample estimator of $D_t$. We use a simulated sample of size $J = 5000$.} The random variable $D_t$ represents the difference between the realized log score and the expected log score under the forecast distribution. In the present example, the expected score is constant (since the forecast distribution is the same in each period), whereas the realized score is a random variable. The figure shows that $D_t$ has a positive mean (marked by the blue line), which is clear evidence against the zero mean condition from \eqref{eqn:entropy}, that is, against auto-calibration. More specifically, the positive mean indicates that the forecast performance in terms of the log score is worse than expected, which is in line with the fact that the forecast variance is too small. The right panel of Figure \ref{fig:ex} visualizes the histogram of (a simulation based estimate of) $U_t$, i.e.\ of the PITs of the log score as defined in \eqref{eqn:PIT_G}.\footnote{Equation (\ref{eqn:pit_ls}) describes the finite sample estimator of $U_t$. We set $J = 5000$.}  The histogram bars tend to increase with $U_t$, displaying a large peak at $U_t > 0.9$. This is clear evidence against the uniformity condition from \eqref{eqn:PIT_G} and thus against auto-calibration. Furthermore, it indicates that the realized scores $g(F_t, Y_t)$ are systematically larger than the scores $g(F_t, X_t)$ simulated under the forecast distribution, again indicating a worse performance than expected.

Of course, in a converse setting where the forecast variance is too large, we would observe a negative mean of $D_t$, as well as an accumulation of small values of $U_t$ (i.e., realized scores would be systematically smaller than scores simulated under the forecast distribution).

\subsection{Implementation of the general tests}

Consider an evaluation sample $\{(F_t,Y_t)\}_{t=1}^T$ consisting of $T$ forecast-observation pairs. The forecast distribution $F_t$ might come in closed form as a CDF $F_t$  itself or as the corresponding density $f_t$.\footnote{Depending on the context we write $F_t$ when referring to the CDF in closed form, but also use $F_t$ to denote the forecast distribution in general and not making explicit in which form it comes, e.g. when denoting a forecast-observation pair as $(F_t,Y_t)$.} The forecast distribution $F_t$ might alternatively by represented by samples $\{X_{j,t}\}_{j=1}^J$, where each $X_{j,t} \in \mathbb{R}^d$ is distributed according to $F_t$. State-of-the-art forecasting methods increasingly take the latter form: For example, meteorological forecasts are often based on an ensemble of draws from a physics-based model \citep[see][for an overview]{VannitsemEtAl2018}. In economics, Bayesian prediction methods using MCMC sampling have become popular in the last decade \citep[e.g.][]{Clark2011}. Simulation-based predictive distributions are particularly relevant in the multivariate case since flexible forms of dependence across variables may not easily be captured by parametric families. In the following, we allow for the case that the simulation samples are dependent across instances $j = 1, \ldots, J$, as is common in MCMC samples, for example. However, we assume that the sequence of draws is strictly stationary and ergodic, with invariant distribution $F_t$. These assumptions are standard and ensure that the simulation output indeed represents the desired forecast distribution \citep{KruegerEtAl2020,Craiu2014}.

If the forecast CDF $F_t$ or PDF $f_t$ is available in closed form, $g(F_t, Y_t)$ is typically straightforward to determine. If the calculation of $g$ requires the forecast distribution in closed form, but the latter is given by a sample $\{X_{j,t}\}_{j=1}^J$, we can, of course, estimate $F_t$ by the empirical CDF and $f_t$ by a kernel density estimator. However, unless the dimension $d$ is small, the curse of dimensionality will make this estimation unreliable, even if we can easily obtain large samples from the forecast distribution. In this case, it is hence preferable to use a dimensionality reduction function that does not require $F_t$ or $f_t$ in closed form. We provide examples below.

To calculate $U_t$ from \eqref{eqn:PIT_G} or $D_t$ from \eqref{eqn:entropy}, the implied forecast distribution $G_t$ of the dimensionality reduction function (evaluated at a specific point, namely $g(F_t,Y_t)$) or its expectation $\E_{F_t} [g(F_t,X_t)]$, respectively, are needed. $G_t$ is usually not available in closed form, even if $F_t$ is. This does not pose a problem as we usually can obtain a large number of draws from $F_t$, i.e.\ a sample $\{X_{j,t}\}_{j=1}^J$. This immediately provides us with a sample from $G_t$, $\{g(F_t,X_{j,t})\}_{j=1}^J$, from which we can calculate the empirical CDF, $\hat{G}_t$, which approximates $G_t$ arbitrarily well with growing $J$ if a Glivenko-Cantelli theorem holds. This is ensured by strict stationarity and ergodicity of $\{X_{j,t}\}$, which carries over to $\{g(F_t,X_{j,t})\}$. This in turn ensures that
\begin{equation*} \label{eqn:Uhat}
\hat{U}_t = PIT_{\hat{G}_t,g(F_t,Y_t)}
\end{equation*}
approximates $U_t$ arbitrarily well, i.e. $\hat{U}_t \stackrel{a.s.}{\rightarrow} U_t$ as $J \rightarrow \infty$. Similarly, a strong law of large numbers (also implied by strict stationarity and ergodicity) for $\{X_{j,t}\}$ ensures that
\begin{equation} \label{eqn:Dhat}
\hat{D}_t= g(F_t,Y_t) - \frac 1 J \sum_{j=1}^J g(F_t, X_{j,t}) \stackrel{a.s.}{\rightarrow} D_t.
\end{equation}

To test the condition from \eqref{eqn:PIT_G}, i.e.\ uniformity of $U_t$, one can use tests of standard uniformity under temporal dependence.\footnote{Temporal dependence of $U_t$ and $D_t$ typically arises in multi-step forecasting.} \cite{Chen2011} shows that many tests proposed in the literature can be regarded as moment-based tests. The class of moment-based tests like \citeauthor{Knueppel2015}'s \citeyearpar{Knueppel2015} test are well-suited for this purpose, since HAC covariance matrix estimators can account for the potential presence of serial correlation in the underlying time series. These time series are transformations of $U_t$ chosen such that their expectation equals the moment of interest, respectively. Therefore, moment-based tests typically only require that a central limit theorem holds for the sample means of these series. To test condition \eqref{eqn:entropy}, we can use a $t$-test with HAC standard errors. Thus, our tests are straightforward to implement (e.g., by using the software accompanying this paper) and have correct asymptotic size under standard assumptions.

If forecasts are generated from an estimated model, the estimation error of the parameters can cause problems in the context of forecast evaluation, as shown by \cite{West1996} for point forecasts. \cite{Chen2011} explains how to address this problem for moment-based calibration tests of density forecasts. However, this approach is quite involved and hardly used. In virtually all applications, density forecasts are evaluated without taking the effects of parameter estimation error into account. This approach is rationalized by \cite{RossiSekhposyan2019} who argue that in this case, the forecast model and the estimation technique are evaluated jointly. See also \cite{GiacominiWhite2006} for a similar argument in the context of relative forecast evaluation.

\section{Score-based calibration tests} \label{sec:score-based_tests}

\subsection{Proper scoring rules as dimensionality reduction functions}

Ideally, dimensionality reduction functions for calibration testing should fulfill four criteria. First, in order to yield a powerful test, $g(F, Y)$ should be sensitive to changes in $F$ and $Y$. This criterion is not trivial as $g$ condenses information from $d$ dimensions into one dimension. A second criterion is applicability: A dimensionality reduction function may lead to powerful tests, but at the same time not be applicable in certain practically relevant situations. A main example here are dimensionality reduction functions that require the forecast distribution in closed form, which are not applicable when the forecasts are given as samples and the dimension is not small enough to make nonparametric estimation of the densities or CDFs feasible. A third criterion is interpretability: In case of a rejection of the null, do the tests, i.e.\ the distribution of $\{\hat{U}_t\}_{t=1}^T$ for the PIT-test or $\frac 1 T \sum_{t=1}^T \hat{D}$ for the $t$-test, provide some constructive information on what is wrong with the forecasts and how they can be improved? A fourth criterion is simplicity regarding the dimensionality reduction functions and the implementation of the associated tests. Practitioners may not be willing to use a test that is overly complicated to understand or implement. Finally, the test results should not depend on the order of the variables to be forecast.\footnote{Note that several tests in the econometric literature do not fulfill this criterion, e.g. those of \cite{Clements-Smith-02} and \cite{Ko-Park-13}.} We argue that proper scoring rules fulfill these five criteria.

Proper scoring rules are the cornerstone of relative forecast evaluation for probabilistic forecasts, where they are used to rank competing forecasts in terms of expected scores. A proper scoring rule $S$ maps a forecast-observation pair $(F,Y)$ to the real line. Propriety is defined as follows \citep{GneitingRaftery2007}:

\begin{Definition}
	A \textbf{scoring rule} $S$ is \textbf{proper} relative to a class of distributions $\mathcal{P}$ if it satisfies
	\begin{equation} \label{eqn:propriety}
	\int S(F_1, y) ~dF_1(y) \le \int S(F_2,y)~dF_1(y)
	\end{equation}
	for all $F_1, F_2$ in $\mathcal{P}$. It is \textbf{strictly proper} if \eqref{eqn:propriety} holds with equality if and only if $F_1 = F_2$.
\end{Definition}

Hence if $Y$ is distributed according to $F_1$, stating $F_1$ yields a smaller (i.e., better) expected score than stating any other distribution $F_2$. In that sense, propriety incentivizes honest and careful forecasting. Strictly proper scores are designed to take into account all facets of a multivariate forecast distribution and the corresponding observation: The optimal forecast is preferred over a forecast that is non-optimal with respect to any single facet. When comparing misspecified forecasts, proper scores should be sensitive to various facets. Thus, proper scoring rules can be expected to perform well with respect to power of the calibration tests, the first criterion mentioned in the previous paragraph.

The most popular proper scoring rules for multivariate probabilistic forecasts are the log score,
\begin{equation*}
	\text{LS}(f,Y) = -\log \left(f(Y) \right),
\end{equation*}
and the energy score,
\begin{equation} \label{es}
	{\text{ES}}(F, Y) = \mathbb{E}_{{F}}||{X}-Y|| - 0.5~\mathbb{E}_{{F}}||{X}-X^*||,	
\end{equation}
where $||\cdot||$ denotes Euclidean distance and ${X}, X^*$ are two independent draws from $F$ \citep{GneitingRaftery2007,SzekelyRizzo2005}.
We use both scores in negative orientation, i.e. smaller scores indicate more accurate forecasts. We will focus on these two specific scores for the rest of this paper due to their popularity and the evidence on their discrimination ability provided in \cite{lerch2017} and \cite{ZielBerk2019}.

Regarding the second criterion, applicability, the log score requires the forecast density $f$ in closed form. While using a kernel density estimate of the forecast density is possible in principle, this approach is challenging even in the univariate setup \citep{KruegerEtAl2020} and thus seems unappealing in the multivariate case. The energy score on the contrary is widely applicable, in that its defining expectations in (\ref{es}) can readily be estimated via empirical averages.

Proper scores as dimensionality reduction functions facilitate the interpretation of test results: In a nutshell, our testing approaches are based on comparing realized scores, $S(F,Y)$, to scores anticipated by the forecaster, $S(F,X)$ with $X \sim F$. If the former tend to be larger than the latter, the forecaster is overconfident in that she anticipates her forecasts to be better than they actually are.\footnote{Of course, the evaluation of confidence is conditional on the score used. A simple example would be a normally distributed target variable and the issuance of a $t$-distributed forecast with correct mean and variance. These forecasts would turn out to be overconfident with respect to the log score, but neither over- or underconfident according to the Dawid-Sebastiani score, because the latter only evaluates the mean and variance forecast.} We detail the aspect of interpretability when discussing the two specific tests below. Finally, our tests based on proper scores are reasonably simple to understand and implement, and their results do not depend on the order of the variables.

Besides their suitability in terms of sensitivity, applicability, interpretability and simplicity, there are further conceptual reasons to use proper scores as dimensionality reduction functions. As noted by \cite{HeldEtAl2010} and \cite{WeiEtAl2017}, the fact that scores are widely used in relative forecast evaluation suggests their usefulness for calibration testing as well. Furthermore, there is a connection between proper scores and calibration, in particular auto-calibration, via the Murphy decomposition of the expected score \citep{Pohle20,gneiting2021,dimitriadis2021}. Finally, our use of scoring rules can be motivated from the perspective of statistical depth functions \citep{zuo2000, mosler2020}. The reason why the classical PIT uniformity result does not generalize to the multivariate case is that quantiles do not readily generalize to the multivariate case due to the lack of an order in $\mathbb{R}^d$ \citep{genest2001}. The univariate PIT basically checks if all quantile levels are hit uniformly. Statistical depth functions, which create a center-outward ordering in $\mathbb{R}^d$, are often proposed as a multivariate generalization of quantiles. A generalization of probabilistic calibration along those lines would thus check if the observations hit all depths of the forecast distribution uniformly. Many depth functions have been proposed \citep[see][]{mosler2020}. \cite{ThorarinsdottirEtAl2016} use the so-called band depth for graphical diagnostic checking of calibration. Using Euclidean depth, perhaps the most basic depth function, as the dimensionality reduction function $g(F,Y)$ in our PIT-based test is equivalent to using the energy score. The same holds true to for the Mahalanobis depth and the Dawid-Sebastiani score.

\subsection{Generalized Box transform test}

When using the PIT-based testing approach with proper scoring rules as dimensionality reduction functions, the PIT of the implied forecast distribution from \eqref{eqn:PIT_G} becomes
\begin{equation} \label{eqn:GBT}
	U_{S} =  \mathbb{P}_F \left( S(F,X) \le S(F,Y) \right)
\end{equation}
with $X \sim F$. For reasons explained in Section \ref{sec:Dimensionality reduction function: existing approaches}, we refer to tests based on $U_S$ as \textit{generalized Box transform (GBT)} tests.

Using the log score, \eqref{eqn:GBT} becomes
\begin{equation} \label{eqn:GBT_LS}
U_{\text{LS}} =  \mathbb{P}_F \left( -\log \left( f(X) \right) \le -\log \left( f(Y) \right) \right).
\end{equation}

As discussed in the previous section for a general $g$, we approximate the PIT of the implied forecast distribution $U_{\text{LS}}$ by drawing a large sample, $\{X_{j,t}\}_{j=1}^J$, from $F_t$ and calculating
\begin{equation}
	\hat U_{\text{LS},t} = \frac{1}{J} \sum_{j=1}^J \mathbf{1} \left( \log \left(f_t \left( X_{t,j} \right) \right) \ge \log \left(f_t \left( Y_{t} \right) \right) \right),
	\label{eqn:pit_ls}
\end{equation}
where $\mathbf{1}(\bullet)$ denotes the indicator function. Subsequently, we test standard uniformity of $\{\hat U_{\text{LS},t}\}_{t=1}^T$ with an appropriate test. We use the test by \cite{Knueppel2015} in our simulations and empirical case studies.

In contrast to the test based on the log score, the GBT test based on the energy score does not require a closed form forecast distribution. Here, $U_S$ from \eqref{eqn:GBT} becomes
\begin{equation*}
U_{\text{ES}} =  \mathbb{P}_F \left( \mathbb{E}_{{F}}||{X} - X^*|| \le \mathbb{E}_{{F}}||{X}-Y|| \right),
\end{equation*}
where $X, {X^*}$ are two independent draws from $F$.\footnote{Using the energy score corresponds to the dimensionality reduction function $
	g(F,Y) = \mathbb{E}_{F}||Y-X||$.
	This function ignores the minuend in (\ref{es}), which drops out when moving to the PIT-based or the $t$-test. This choice of $g$ also establishes the connection between the use of the energy score and the Euclidean depth function mentioned above.}
In order to estimate $U_{\text{ES}}$, we use two distinct samples $\{X_{t, i}\}_{i=1}^{J_0}$ and $\{X^*_{t,j}\}_{j=1}^{J_1}$, both of which are drawn from $F_t$, and set
\begin{equation}
\hat U_{\text{ES},t} = \frac{1}{J_1}\sum_{j=1}^{J_1}\mathbf{1} \left\{\frac{1}{J_0}\sum_{i=1}^{J_0}||X_{t,i}-X^*_{t,j}|| \le \frac{1}{J_0}\sum_{i=1}^{J_0}||X_{t,i} - Y_t|| \right\}.
\label{eqn:pit_es}
\end{equation}
We then test standard uniformity of $\{\hat U_{\text{ES},t}\}_{t=1}^T$. In practice, it may sometimes be more convenient to use one sample and split it in the middle to obtain two samples than drawing two separate independent samples. This appears unproblematic if the dependence between the two samples is sufficiently weak.\footnote{Alternatively, the following estimator based on a single sample $\{X_{t,j}\}_{j=1}^{J}$ could be used:
	$$\widetilde U_{\text{ES}, t} = \frac{1}{J} \sum_{j=1}^{J} \mathbf{1}\left(\frac{1}{J}\sum_{\substack{i=1}}^{J} ||X_{t,i}-X_{t,j}|| \le \frac{1}{J}\sum_{i=1}^{J} ||X_{t,i}-Y_t||\right).$$
	Simulation results suggest that this estimator performs similarly to the estimator in \eqref{eqn:pit_es}, but we use the latter throughout this paper.\label{footnote:exact_estimator}}

To interpret the test results or to check calibration more informally in a graphical manner, histograms of $\{\hat U_{\text{ES}, t}\}_{t=1}^T$ can be used in a similar way as PIT histograms are used in the univariate case. If the height of the histogram bars increases over $[0,1]$ as in the example in the previous section, the forecaster is overconfident as the realized scores tend to be higher than the ones expected by the forecaster. Vice versa, if the histogram decreases, the forecaster is underconfident. Further shapes of the histograms may also hint to certain deficiencies of the forecasts: For example a U-shape means that the forecaster usually tends to give an assessment of expected forecast performance in a medium range, while the true forecast performance often tends to be much better or much worse. Thus, while they may not be over- or underconfident, the forecasts have difficulties in distinguishing between situations of good and bad expected forecast performance (i.e., low and high expected scores) and are not that informative in that respect. Reversely, a hump-shaped histogram arises because the assessments of expected forecast performance tend to be more extreme than the true forecast performance.

\subsection{Entropy test}

Using proper scores as dimensionality reduction functions in our $t$-test approach based on \eqref{eqn:entropy} amounts to a comparison of the forecaster's expected score, $\mathbb{E}_F [S(F,X)]$, and the realized score, $S(F,Y)$, via their difference, $D=S(F,Y) - \mathbb{E}_F [S(F,X)]$. $S(F, Y)$ represents the `ex post' performance of the forecast distribution $F$, with larger values corresponding to worse performance and higher prediction uncertainty of $Y$. The `ex ante' expected score under $F$, $\mathbb{E}_F [S(F,X)]$, is known as the generalized entropy of $F$ (see e.g.\ \citealp{GneitingRaftery2007}), which is why we call this testing approach the \textit{entropy test}. Related comparisons between ex post (or `objective') uncertainty and ex ante (or `subjective') uncertainty  have been considered for univariate forecast distributions by \cite{galvao2019}, \cite{Clements2014} and \cite{KruegerPavlova2020} in the context of economic survey forecasts. The entropy test is easily interpretable: If $E[D]>0$, realized uncertainty exceeds expected uncertainty, i.e. the forecaster is overconfident in an 'average' sense. The opposite holds if $E[D]<0$. Further, plotting the difference in the two uncertainties, $D$, over time can yield additional insights beyond considering its mean only: This diagnostic plot may reveal periods of over- and underconfidence and thus point to deficiencies in a forecaster's ability to quantify uncertainty in certain situations.

For the log score, $D$ from equation \eqref{eqn:entropy} becomes
$$D_{\text{LS}} = - \log \left( f(Y) \right) - \E_F [- \log \left( f(X) \right)].$$
Thus, here the entropy test amounts to using a sample $\{X_{j,t}\}_{j=1}^J$ from $F_t$ to compute
\begin{equation}
\hat{D}_{\text{LS},t} = - \log \left( f_t (Y_t) \right) + \frac 1 J \sum_{j=1}^J  \log \left( f_t(X_{j,t}) \right)
\label{eqn:dhat_logscore}
\end{equation}
and then to perform a $t$-test with HAC standard errors for the null $E[D_t]=0$ on $\{\hat{D}_{\text{LS},t}\}_{t=1}^T$.

For the energy score $D$ is given by
$$D_{\text{ES}} = \E_{{F}} ||{X}-Y|| - \mathbb{E}_{F}||{X}-X^*||.$$

In line with equation (\ref{eqn:pit_es}), this term can be estimated based on two distinct samples $\{X_{t, i}\}_{i=1}^{J_0}$ and $\{X_{t,j}^*\}_{j=1}^{J_1}$:
\begin{equation}
\hat D_{\text{ES},t} =
\frac{1}{J_0}\sum_{i=1}^{J_0}||X_{t,i} - Y_t|| - \frac{1}{J_0J_1}
\sum_{i=1}^{J_0}\sum_{j=1}^{J_1}||X_{t,i}-X_{t,j}^*||.\label{eqn:dhat_es}
\end{equation}
Alternatively, Equation 7 of \cite{GneitingEtAl2008} suggests to approximate $D_t$ by a single sample $\{X_{j,t}\}_{j=1}^J$ from $F_t$ by setting
\begin{equation}
\widetilde D_{\text{ES},t} = \frac{1}{J}\sum_{j=1}^J ||X_{t,j}-Y_t|| - \frac{1}{J^2}\sum_{i=1}^J \sum_{j = 1}^J ||X_{t,i}-X_{t,j}||.\label{eqn:dhatfull}
\end{equation}
The latter estimator is the analogue to the PIT estimator in Footnote \ref{footnote:exact_estimator}.

\subsection{Dimensionality reduction functions of existing approaches} \label{sec:Dimensionality reduction function: existing approaches}
Many of the existing tests are special cases of our PIT-based approach: They first apply a dimensionality reduction function and then test the condition from \eqref{eqn:PIT_G}, i.e.\ standard uniformity of the PIT of the implied forecast distribution. We now discuss dimensionality reduction functions used in some of these tests, noting that there are more functions used especially in meteorology (e.g.\ \citealp{thorarinsdottir2018}; \citealp{wilks2017}).

\citet{DovernManner2020} propose two order-invariant tests based on the decomposition of $F$ into conditional CDFs and the corresponding conditional PITs. Their preferred test uses the dimensionality reduction function
\begin{equation} \label{eqn:DoMa}
g_{DM}(F,Y) = \sum_{i=1}^d \left[ \Phi^{-1} \left( PIT_{F_{Y^{(i)}|Y^{(1)},..., Y^{(i-1)},Y^{(i+1)},...,Y^{(d)}},Y^{(i)}} \right) \right]^2,
\end{equation}
where $\Phi^{-1}$ is the quantile function of the standard normal distribution and $F_{Y^{(i)}|\dots}$ denotes the CDF of $Y(i)$ conditional on all other variables. The applicability of this test is limited to cases where the forecast distribution is given in closed form or where the dimension $d$ is small enough such that a reliable nonparametric density estimation is feasible.\footnote{Under the assumption that $Y$ is multivariate normal, $G$, i.e.\ the distribution of of $g_{DM}(F,X)$ under the null hypothesis, is known and $U$ can be calculated directly without using draws from the forecast distribution.}

The dimensionality reduction functions showing up in the meteorological literature, where they are rather used for graphical diagnostic checking than for hypothesis testing, do not use conditional PITs. Nevertheless, most of them employ the PIT in some form. The multivariate rank histogram or Copula PIT \citep{GneitingEtAl2008, ZiegelGneiting2014} uses the multivariate PIT as dimensionality reduction function:
\begin{equation*} \label{eqn:CoPIT}
g_{CoPIT}(F,Y) = F(Y).
\end{equation*}
However, as the $\leq$-relation on $\mathbb{R}^d$ only provides a partial order,
the event $\{ X \le Y \}$ is very uninformative even for moderate dimensions $d$. Hence $F(Y) = \mathbb{P}_F(X \le Y)$ will often be close to zero, making it hard to distinguish between an auto-calibrated forecast distribution $F$ and miscalibrated alternatives.\footnote{Note that here we compactly write $\mathbb{P}_F(X \le Y)$ for $\mathbb{P}(X \le Y|F)$ as well.} This leads to tests based on the Copula PIT having low power, a problem getting worse with increasing $d$.

A simple approach proposed by \cite{ThorarinsdottirEtAl2016} is the average rank histogram or average PIT, which just averages over all $d$ marginal PITs:
\begin{equation}
g_{AvPIT}(F,Y) = \frac 1 d \sum_{i=1}^d F^{(i)} \left( Y^{(i)} \right).
\label{eqn:AvPIT}
\end{equation}
This approach does not explicitly employ information contained in $F$ beyond its marginals. We consider it in our simulations in the next section and shed some light on its usefulness there.

Remarkably, the Box transform as introduced by \cite{GneitingEtAl2008} for checking calibration graphically does not use the PIT in the dimensionality reduction function $g(F, Y)$. It is defined as
\begin{equation} \label{eqn:BoxTransform}
U_{BT}=1 - \mathbb{P}_F \left(f (X) \le f (Y) \right),
\end{equation}
where $f$ is the forecast density corresponding to $F$, and $X \sim F$. Note that $U_{BT}$ is equivalent to $U_{LS}$ from \eqref{eqn:GBT_LS}, since the logarithm is a monotonic function. Thus, our PIT-based test with a proper score as a dimensionality reduction function can be viewed as a generalization of testing with the Box transform, hence the name `generalized Box transform (GBT) test'.

In our simulations, we consider the Dovern-Manner test and a test based on the average PIT. Concerning the four criteria mentioned above, the Dovern-Manner test will turn out to be appealing in terms of power, but its applicability is limited to the case of closed-form or low-dimensional forecast distributions. Furthermore, the test is hard to interpret (as the dimensionality reduction function from \eqref{eqn:DoMa} averages over many terms involving conditional PITs) and to implement. In contrast to the Dovern-Manner test, tests based on the average PIT have very limited power in our simulations, but are relatively simple to interpret and easy to implement. Using the copula PIT as a dimensionality reduction function seems difficult to motivate in terms of either criterion.

\section{Simulation studies} \label{sec:simulations}
\label{sec:sim_baseline}

We next compare various tests via simulation, closely following the design of \citet[Section 3.1]{DovernManner2020}. The null hypothesis $H_0$ states that the data follows a multivariate normal distribution where each of the $d$ variables has a mean equal to 0, a variance equal to 1, and its correlation with each other variable equals $0.5$. Simulating data from this distribution for different dimensions $d$ and different numbers of periods $T$ will inform us about the size of the tests considered. To evaluate the tests' power, we use the multivariate forecast distribution of $H_0$, and data generated by four alternative distributions:
\begin{itemize}
	\item $H_1$: Multivariate normal distribution, variance of each variable equals $1.1^2$, all other parameters (i.e., means and correlations) like under $H_0$.
	\item $H_2$: Multivariate normal distribution, correlation of each pair of variables equals 0.4, all other parameters (i.e., means and variances) like under $H_0$.
	\item $H_3$: Multivariate $t$ distribution with 8 degrees of freedom, rescaled such that covariance is like under $H_0$.
	\item $H_4$: Multivariate Gaussian constant conditional correlation (CCC)-GARCH(1,1) model of \cite{Bollerslev1990} with GARCH parameters $\omega=0.05, \alpha=0.1, \beta=0.85 $; means and unconditional covariance matrix like under $H_0$.
\end{itemize}
Note that our setup only differs from \cite{DovernManner2020} with respect to $H_3$, where we rescale the distribution such that differences from $H_0$ only occur in fourth (and larger even) moments. This makes it easier to distinguish the results from those under $H_1$, where `under $H_i$' means that the forecaster uses the distribution of $H_0$, while the data-generating process (DGP) is given by $H_i$.

We only consider the $Z_t^{2\dagger}$ test of \cite{DovernManner2020}, since applying their $Z_t^{2*}$ test would be too time-consuming even in the case of normal forecast densities unless $d$ is small. Moreover, both tests have similar size and power properties. We also employ the test based on the average-rank histogram proposed by \cite{ThorarinsdottirEtAl2016}.
For the entropy tests, the difference $\hat{D}_t$ between the realized score and the expected score is regressed on a constant, and the null hypothesis of this constant being equal to zero is tested. $\hat{D}_t=\hat{D}_{ES,t}$ is described in equation \eqref{eqn:dhat_es} for the case of the energy score. For the case of the log score, $\hat{D}_t=\hat{D}_{LS,t}$ is given in \eqref{eqn:dhat_logscore}. For the tests based on the generalized Box transform (GBT), we use Kn\"uppel's test.\footnote{\cite{DovernManner2020} employ \citeauthor{Neyman1937}'s \citeyearpar{Neyman1937} smooth test, but this test is hardly applied in the corresponding literature. In the simulation design of \cite{DovernManner2020}, \citeauthor{Neyman1937}'s \citeyearpar{Neyman1937} smooth test has more power than Kn\"uppel's test, but the opposite would hold if, for instance, the true variances would be smaller than 1 under $H_1$.} The PITs entering the test are given by \eqref{eqn:pit_es} for the energy score and by \eqref{eqn:pit_ls} for the log score. As implied by \eqref{eqn:AvPIT}, the average rank PITs are calculated as
\begin{equation*}
\hat U_{Av,t} = \frac{1}{J} \sum_{j=1}^J \mathbf{1}\left(\frac{1}{d}\sum_{i=1}^{d} F \left( X_{t,j}^{(i)} \right) < \frac{1}{d}\sum_{i=1}^{d} F \left( Y_{t}^{(i)} \right)\right).
\end{equation*}

The results in Table \ref{tab:DovMan} show that all tests considered have good size properties. With respect to power, however, pronounced differences arise. The Dovern-Manner test and the log-score-based GBT test tend to attain relatively high power in all situations considered, and their power increases with the number of variables $d$ and the number of periods $T$. Power also increases with $d$ and $T$ for the energy-score-based GBT test. The power of the average rank test increases with $T$, but often remains fairly constant or even decreases when $d$ increases. This behavior can occur if the individual ranks have the same correlation under the null and the alternative hypothesis, but different variances. These conditions turn out to hold (at least approximately) under $H_1$, $H_3$ and $H_4$.\footnote{
	In all situations considered, the mean of the average rank equals 0.5. The variance of the average rank is given by $\sigma^2(\rho + \frac{1-\rho}{d})$, where $\sigma^2$ denotes the variance of an individual rank and $\rho$ denotes the correlation between two distinct individual ranks. This simple formula holds because in each DGP used, $\sigma^2$ and $\rho$ are constant across all individual ranks and pairs of distinct individual ranks, respectively. Letting $\sigma_{H_0}^2$ and $\rho_{H_0}$ denote these quantities under $H_0$, and letting $\sigma_{H_1}^2$ and $\rho_{H_1}$ denote these quantities under $H_1$, two interesting special cases occur. If $\rho_{H_1} = \rho_{H_0} = \bar{\rho}$, the variances of the average ranks differ by $(\bar{\rho}+\frac{1-\bar{\rho}}{d})(\sigma_{H_1}^2-\sigma_{H_0}^2)$. Obviously, this difference \emph{decreases} (in absolute terms) with the number of variables $d$, as observed in the cases $H_1$ and $H_4$. On the other hand, if $\sigma_{H_1}^2 = \sigma_{H_0}^2 = \bar{\sigma}^2$, the difference becomes $\bar{\sigma}^2 (1-\frac{1}{d})(\rho_{H_1} - \rho_{H_0})$, which \emph{increases} (in absolute terms) with the number of variables $d$, as observed in the case $H_2$. While these mechanisms are important to understand the behavior of the average rank test, they cannot explain every aspect, since higher moments like kurtosis also play a role.
}
The entropy tests can also suffer from a lack of power, but at least in the case of the log score under $H_3$, this result has a simple explanation. The expected score in this case only depends on the mean vector and the covariance matrix of the forecast distribution. Since the true distribution has the same mean vector and  covariance matrix, the expected score and the average realized score are identical.\footnote{Note that the same result would be obtained using the score proposed by \cite{DawidSebastiani1999}, which considers the first two moments only. } Actually, using the log score and a normal forecast distribution under $H_0$ and $H_3$, conducting the entropy test amounts to measuring the difference of the respective averages of the Mahalanobis distance from zero in this setup. The energy score is based on the Euclidean distance instead. The low power of the energy-score-based entropy test under $H_3$ thus indicates that the expected Euclidean distance under $H_0$ and the average realized Euclidean distance are similar. The same holds under $H_1$. Under $H_2$, due to the larger variance of the DGP, the expected distance is obviously smaller than the average realized distance, be it of the Mahalanobis or the Euclidean type. Accordingly, both entropy tests have relatively high power in this case.

Summing up, in the simulation setup based on \cite{DovernManner2020}, the GBT tests turn out to be powerful tools to assess the calibration of multivariate forecast densities. The entropy tests do not have power against certain alternatives, but their results are easier to interpret, as done above for their behavior under $H_1$, $H_2$ and $H_3$.
Finally, it is worth stressing that all tests except for the Dovern-Manner test are fairly easy to apply to non-Gaussian distributions. If the forecast density is unknown or hard to approximate, but a sample from this distribution is available, only the average rank test and the energy-score-based tests can be used.

Additional simulation results for the GBT and entropy test covering time series dependence, $t$-distributed forecast distributions and heteroskedasticity can be found in Appendix \ref{sec:appendix}.

\begin{table}
	\begin{small}
		\centering
		\input{./dmcompare6}
		\caption{Size and power of tests for multivariate calibration}
		\label{tab:DovMan}
	\end{small}
\end{table}

\section{Macroeconomic case study: Forecasting US growth, inflation, and interest rates} \label{sec:applications}

We next apply our calibration tests to forecasts of three quarterly US macroeconomic variables: The gross domestic product (GDP) growth rate, the inflation rate as measured by the GDP deflator, and the three-month interest rate (T-Bill rate).\footnote{For GDP and the GDP deflator, we use real-time data as provided by the Philadelphia Fed (series codes \textsf{ROUTPUTQvQd} and \textsf{PQvQd}). We consider logarithmic growth rates, and define the outcome as the second available data vintage. For the interest rate, we consider un-revised data from the FRED data base (series code \textsf{DTB3}).} We consider forecast horizons of $h = 0, 1, \ldots, 4$ quarters ahead, and our evaluation sample ranges from 1965:Q4 to 2021:Q3 (target dates).

We employ a time-varying vector autoregressive model as proposed by \cite{Primiceri2005} and \cite{Del2015} to generate forecasts, which is in line with \cite{DovernManner2020}.\footnote{The corresponding application of their tests requires challenging nonparametric kernel estimation procedures plagued by the curse of dimensionality.} The model is estimated via Bayesian methods, where we follow the default implementation in the bvarsv package for R \citep{Krueger2015}.\footnote{A description of the package is available at \url{https://github.com/FK83/bvarsv/blob/master/bvarsv_Nov2015_website.pdf}.} Briefly, the model is given by
\begin{eqnarray*}
	Y_t &=& \nu_t + A_t Y_{t-1} + \varepsilon_t\\
	\varepsilon_t &=& \mathcal{N}(0, \Sigma_t),
\end{eqnarray*}
where $Y_t$ is the trivariate vector covering GDP growth, inflation and the interest rate in quarter $t$, $\varepsilon_t$ is a trivariate vector of error terms, and $\mathcal{N}$ denotes the trivariate normal distribution. The model allows the parameters in $\nu_t, A_t$ and $\Sigma_t$ to vary over time, following random walk type specifications. Time variation in $\Sigma_t$ is especially important empirically, as it accounts for time-varying volatility in macroeconomic time series. The prior choices used for Bayesian estimation seek to balance flexibility with stability by discouraging excessive time variation in the model parameters. For comparison, we also consider a constant-parameter variant of the model, in which we choose very tight priors that `shut off' time variation in $\eta_t, A_t$ and $\Sigma_t$. All other prior choices are the same.

At each forecast date, we produce $100\,000$ draws from the model's predictive distribution. We retain every tenth of these draws and split the resulting sequence in two halfs, thus resulting in two sequences of $5000$ draws each. We implement our proposed calibration tests based on these sequences, using the split-sample estimator described in Sections 3.2 and 3.3.
We further obtain an analytical forecast density as required for the logarithmic score by exploiting the mixture-of-parameters form implied by the Bayesian predictive distribution \citep[cf.][]{KruegerEtAl2020}.

\begin{table}
	\centering
	\begin{tabular}{cc}
		\multicolumn{2}{c}{\textbf{Stochastic volatility model (Primicieri)}} \\
		Pre-Covid sample (1965:Q4 - 2019:Q4) & Full sample (1965:Q4 - 2021:Q3)  \\
		
		\begin{tabular}{rrrrr}
			\toprule
			$h$ & ES$_{\text{D}}$ & ES$_{\text{GBT}}$ & LS$_{\text{D}}$ & LS$_{\text{GBT}}$\\
			\midrule
0 & 17.51 & 40.03 & 21.91 & 52.65\\
1 & 22.05 & 38.01 & 24.08 & 16.49\\
2 & 23.33 & 34.61 & 23.62 & 75.54\\
3 & 41.00 & 65.51 & 27.60 & 32.67\\
4 & 67.88 & 84.48 & 11.84 & 6.83\\
			\bottomrule
		\end{tabular} &
		\begin{tabular}{rrrrr}
			\toprule
			$h$ & ES$_{\text{D}}$ & ES$_{\text{GBT}}$ & LS$_{\text{D}}$ & LS$_{\text{GBT}}$\\
			\midrule
0 & 74.10 & 27.51 & 15.24 & 33.69\\
1 & 83.05 & 66.59 & 12.90 & 17.32\\
2 & 85.15 & 43.49 & 11.98 & 50.71\\
3 & 71.91 & 86.86 & 12.22 & 17.87\\
4 & 53.46 & 87.16 & 5.65 & 2.99\\
			\bottomrule
		\end{tabular} \\
	&\\
	&\\
	
	\multicolumn{2}{c}{\textbf{Constant parameter model}} \\
	Pre-Covid sample (1965:Q4 - 2019:Q4) & Full sample (1965:Q4 - 2021:Q3)  \\
	
	\begin{tabular}{rrrrr}
		\toprule
		$h$ & ES$_{\text{D}}$ & ES$_{\text{GBT}}$ & LS$_{\text{D}}$ & LS$_{\text{GBT}}$\\
		\midrule
0 & 5.15 & 0.04 & 88.56 & 0.00\\
1 & 13.60 & 0.24 & 97.78 & 0.01\\
2 & 20.91 & 0.06 & 97.47 & 0.05\\
3 & 30.86 & 0.18 & 94.83 & 0.05\\
4 & 48.88 & 1.20 & 75.42 & 0.26\\
		\bottomrule
	\end{tabular} &
	\begin{tabular}{rrrrr}
		\toprule
		$h$ & ES$_{\text{D}}$ & ES$_{\text{GBT}}$ & LS$_{\text{D}}$ & LS$_{\text{GBT}}$\\
		\midrule
0 & 85.48 & 0.03 & 36.83 & 0.00\\
1 & 94.41 & 0.16 & 36.90 & 0.01\\
2 & 97.77 & 0.05 & 33.72 & 0.02\\
3 & 93.13 & 0.23 & 32.50 & 0.02\\
4 & 79.67 & 0.67 & 25.70 & 0.18\\		\bottomrule
	\end{tabular}
	
	\end{tabular}
	\caption{Test results ($p$-values in percent) for trivariate forecast distributions of US GDP growth, inflation, and TBILL rate. $h$ denotes forecast horizon in quarters. Columns refer to test variants. `ES' and `LS' denote the energy and logarithmic score. `D' and `GBT' refer to entropy difference and generalized Box transform. \label{tab1}}
\end{table}

Table \ref{tab1} presents the results of our calibration tests. The left panel refers to a sample period ending in 2019:Q4, excluding the very large Covid-related outliers at the end of the sample period.\footnote{In particular, the annualized GDP growth rates in in 2020:Q2 and 2020:Q3 equal -37.7\% and +28.8\%.} The test results provide no evidence against auto-calibration of the Primiceri model, with all $p$-values exceeding 5\%. For the full evaluation sample including the recent pandemic period (right panel), some of the $p$-values of the tests based on the logarithmic score decrease, hence providing more evidence against auto-calibration. That said, the case against auto-calibration is still weak overall. The bottom panels of Table \ref{tab1} refer to the constant parameter model. In both sample periods (bottom left and bottom right panels), the GBT version of the test clearly rejects auto-calibration, with all $p$-values being smaller than $1.2 \%$. By contrast, the entropy version of the test does not reject auto-calibration, with all $p$-values exceeding $5\%$. We return to this discrepancy below.

\begin{figure}[h]
	
	\centering
	\begin{tabular}{cc}
		\multicolumn{2}{c}{\textbf{~~~~~Stochastic volatility model (Primiceri)}}  \\ 
		\includegraphics[width=.45\textwidth]{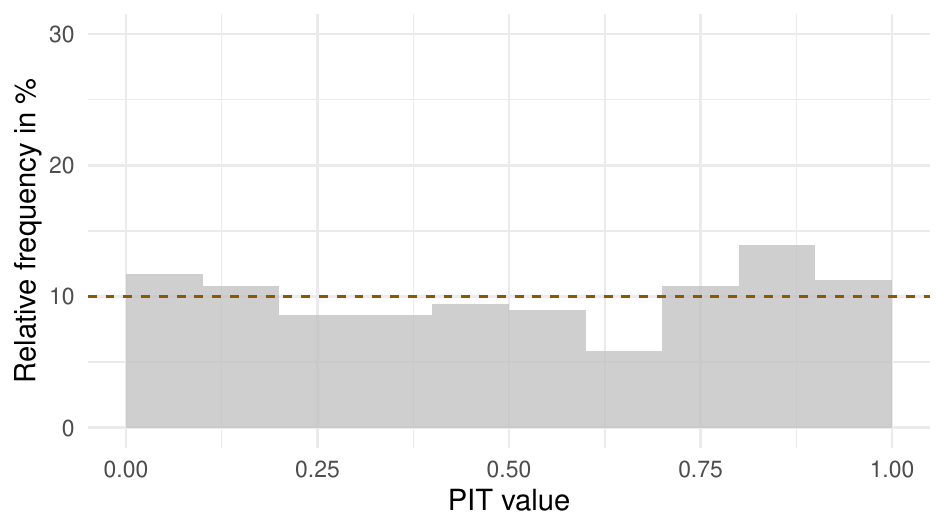} &
		\includegraphics[width=.45\textwidth]{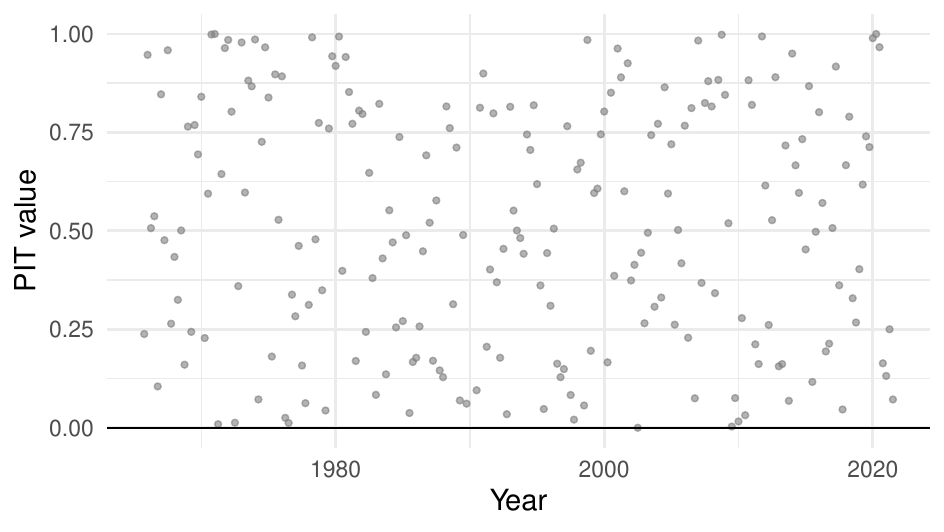}  \\[.5cm]
		\multicolumn{2}{c}{\textbf{~~~~~ Constant parameter model}} \\		
		\includegraphics[width=.45\textwidth]{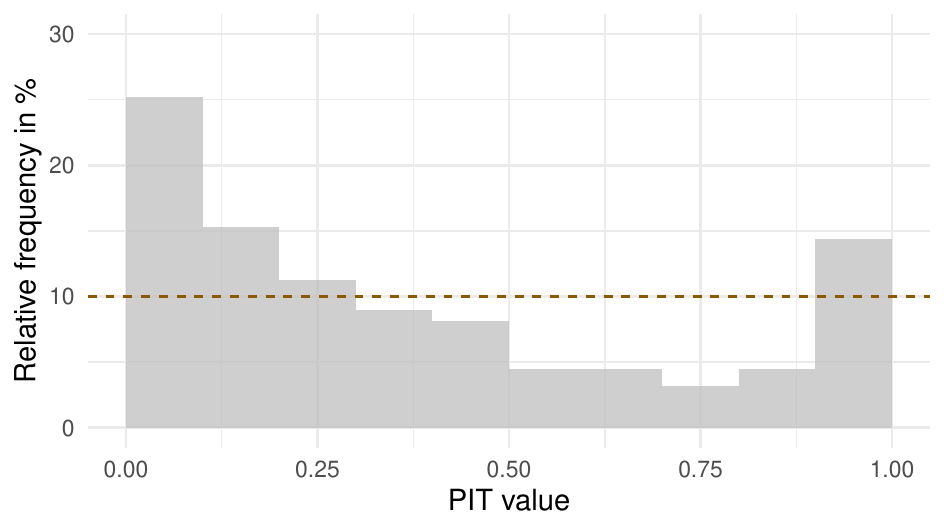} &
		\includegraphics[width=.45\textwidth]{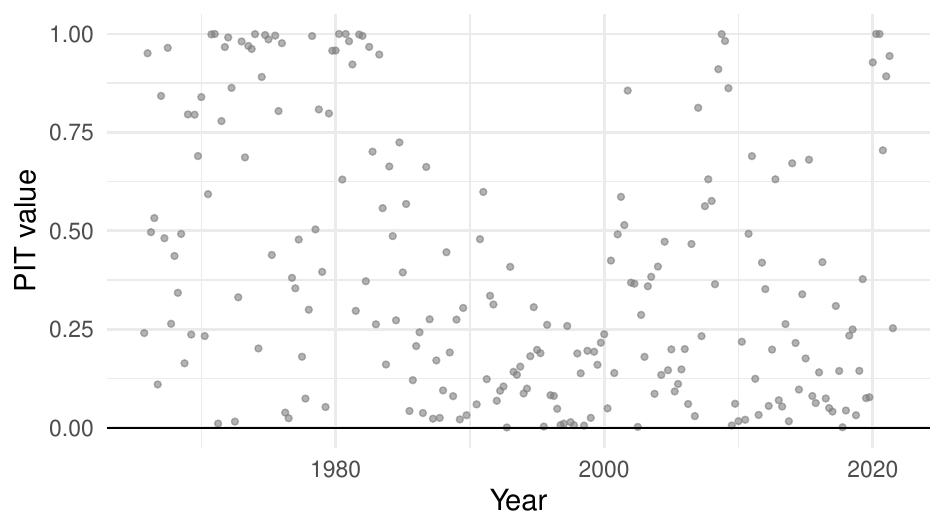} 
		\\
	\end{tabular}
	\caption{Score PIT histograms and time series plots of score PITs ($\hat U_{\text{LS},t}$) for the log score for the macroeconomic case study at $h=0$. 	\label{fig:Primiceri_PIT}}
\end{figure}

We also propose diagnostic plots that can accompany our tests or even be used without conducting a test. They can provide valuable insights into the problems of a forecasting approach and hint to directions for improvement. We present them for the log score and forecast horizon $h = 0$ for brevity. The results for the energy score and for other forecast horizons are qualitatively similar. The left hand side of Figure \ref{fig:Primiceri_PIT} presents score PIT histograms (based on $\hat U_{\text{LS},t}$ from Equation \ref{eqn:pit_ls}) for both models. For the Primiceri model the histogram is very close to uniform, which is in line with the large $p$-value (see top panel of Table \ref{tab1}). For the constant parameter model the histogram has an asymmetric U-shape, where mainly very low, but also very high score values arose more often than anticipated by the forecast distributions and central to moderately high score values much less often. That is, the forecast distributions often tended to overestimate uncertainty, sometimes underestimated it and too rarely predicted a medium level of uncertainty. The plots on the right hand side give further insights into the reasons for these histogram shapes and a very interesting assessment of the performance of the models over time. There we plot the score PITs (again $\hat U_{\text{LS},t}$ from Equation \ref{eqn:pit_ls}) over time. For the constant parameter model the realized scores very often fell into the far right of the implied forecast distribution for the scores prior to the mid-1980s, meaning that the model underestimated the uncertainty in that period. During the Great Moderation and also between the Global Financial Crisis and Covid-19 the vast majority of realized scores fell into the lower half of the anticipated forecast distribution for the scores, meaning that the model severely overestimated uncertainty for those periods. The Global Financial Crisis and the Covid-19 period are visible via PITs close to 1. We will discuss the latter period in more detail below. In contrast, the stochastic volatility model seems to adapt remarkably well to changing uncertainty, exhibiting hardly any patterns in the PITs over time. 

\begin{figure}[h]
	 
	\centering
\begin{tabular}{c}
	\textbf{~~~~~ Stochastic volatility model (Primiceri)} \\
	\includegraphics[width = .5\textwidth]{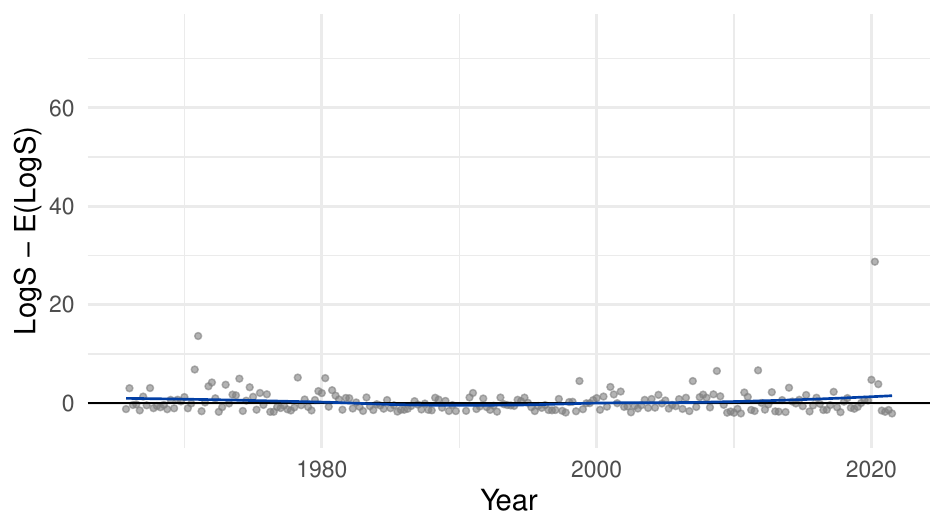} \\[.5cm]
	\textbf{~~~~~ Constant parameter model} \\
		\includegraphics[width = .5\textwidth]{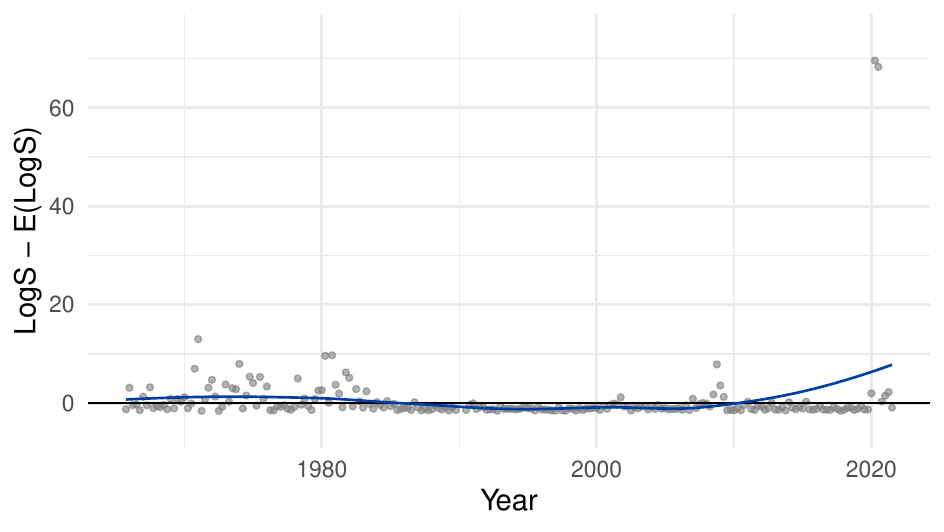}
	\\
\end{tabular}
	\caption{Time series plot of difference between realized and expected log scores ($\hat D_{\text{LS},t}$) for the macroeconomic case study at $h=0$. Blue line shows a nonparametric time trend, based on local quadratic regression as implemented in the R function \textsf{loess}. \label{fig:Primiceri}}
\end{figure}

Also the difference between realized and expected scores, which is central to our entropy test, can offer further insights. The top panel of Figure \ref{fig:Primiceri} plots this difference over time for the Primiceri model, again focusing on the log score (i.e., considering $\hat D_{\text{LS},t}$ defined in Equation \ref{eqn:dhat_logscore}) and forecast horizon $h = 0$. Recall that large values of $\hat D_{\text{LS},t}$ correspond to overconfidence (actual performance is worse than expected performance), and vice versa for small values of $\hat D_{\text{LS},t}$. The plot shows no clear tendency towards either over- or underconfidence before 2020. The Covid-related outlier in 2020:Q2 corresponds to a very large value of $\hat D_{\text{LS},t}$, indicating a massive forecast error that was essentially ruled out by the model. By contrast, the last few values for $\hat D_{\text{LS},t}$ are among the smallest in the sample, indicating  better-than-expected forecast performance. This result can be explained by the model's drastic upward correction of forecast variance in response to the 2020:Q2 and 2020:Q3 observations, resulting in a forecast variance that seems implausibly high. This issue and the related discussion of how to treat Covid-19 outliers has received considerable attention recently \citep[e.g.][]{CarrieroEtAl2021}. 

The bottom panel of Figure \ref{fig:Primiceri} shows an analogous plot for the constant parameter model. In addition to the outlier in 2020:Q2, the bottom panel features an additional outlier in 2020:Q3. This effect arises because the constant parameter model does not strongly revise its volatility prediction upward after 2020:Q2, and is hence surprised by the very high positive growth rate in 2020:Q3 reflecting a growth rebound effect. Note also that the larger value for $\hat D_{\text{LS}_t}$ in 2020:Q2 (about $70$ in bottom panel, compared to $30$ in top panel) arises because the constant parameter model features less probability mass in the far left tail of the forecast distribution, thus producing an even larger log score.

The bottom panel of Figure \ref{fig:Primiceri} also illustrates why the entropy test does not reject auto-calibration of the constant parameter model: Periods of over-confidence (such as before the Great Moderation) and periods of under-confidence (such as in the 1990s) offset each other, such that the hypothesis that $\mathbb{E}(\hat D_{\text{LS}_t}) = 0$ cannot be rejected. While this `blind spot' may be viewed as a disadvantage of the entropy test, it reflects the test's simplicity which in turn enables clear feedback for model fitting. Adjustments of the entropy test which increase power at the cost of adding complexity (e.g., an extension in the spirit of the fluctuation test of \citealp{GiacominiRossi2010}) are possible but beyond the scope of this paper.

\section{Finance case study: Forecasting international stock returns} \label{sec:application_finance}

As a second case study, we consider predicting the joint distribution of ten international stock indexes from Brazil (BSVP), France (FCHI),  Germany (GDAXI), Hong Kong (HSI), Japan (N225), the UK (FTSE), the US (DJI), Spain (IBEX), Switzerland (SSMI), and Europe as a whole (STOXX50E). The data was kindly provided by the Oxford-Man realized library.

We consider forecasts for each business day from 2010 to 2021, based on a rolling window of $1000$ business days. In addition to daily (logarithmic, close-to-close) returns, we use realized kernel measures (\citealp{Barndorff2008,Barndorff2009}) included in the data set. We delete days with missing return or regressor observations for either of the indexes, resulting in an evaluation sample of $2\,496$ observations.

Predicting high-dimensional return distributions is challenging: First, financial return data may feature stylized facts such as conditional heteroskedasticity or skewness that should be addressed for each indidual index. Second, the curse of dimensionality precludes overly complex multivariate distribution models. Here we consider a pragmatic approach that seeks to address this situation:
\begin{enumerate}
	\item For each index $k$, we fit a quantile regression model of the form $$q_\tau(R_t^k|\mathcal{I}_{t-1}) = \beta_0 + \beta_1 ~\text{RK}_{t-1}^k + \beta_2 \times \frac{1}{5} \sum_{l=1}^5 \text{RK}_{t-l}^k
	+ \beta_3 \times \frac{1}{22} \sum_{l=1}^{22} \text{RK}_{t-l}^k,$$
	where $q_\tau(R_t^k|\mathcal{I}_{t-1})$ denotes the $\tau$-quantile of the return $R_t^k$ given $\mathcal{I}_{t-1}$, and $\text{RK}_{t-l}^k$ denotes the realized kernel measure for index $k$ on day $t-l$. We consider a dense grid of quantile levels $\tau = 1/1001, \ldots, 1000/1001$.
	\item Based on the quantile regressions, we consider a spread measure $S_t^k = \hat q_{.6915}(R_t^k|\mathcal{I}_{t-1})-  \hat q_{.3085}(R_t^k|\mathcal{I}_{t-1})$, i.e.\ the difference between the predicted quantiles at levels $69.15\%$ and $30.85\%$. These levels are chosen such that they correspond to a spread measure of one for a standard normal distribution, thus mimicking the standard deviation.
	\item For each index $k = 1, \ldots, 10$ and day $t = 1, \ldots, T$ in the training sample, we then compute the standardized return $Z_t^k = R_t^k/S_t^k$.
	\item We then form a forecast distribution for date $T+1$ by pairing the quantile regression based univariate distributions according to the empirical (rank) copula prescribed by the standardized returns $\{Z_t^k = R_t^k/S_t^k\}_{t,k}$. This step finally yields a forecast sample of $1000$ draws for the $10$-variate return vector of interest.
\end{enumerate}

The idea to use quantile regression at various levels for predicting a distribution has been used by \cite{AdrianEtAl2019}, among others. A distinct feature of our approach is that we consider a dense grid of $1000$ quantile levels. We then sort the resulting predictions in order to obtain a monotone quantile curve \citep[see][]{ChernozhukovEtAl2010}. Fitting a parametric curve to the quantile regression output \citep[as in][]{AdrianEtAl2019} is not necessary in our case since our evaluation methodology based on the energy score does not require a forecast density.

While simple to implement via the R package \verb|quantreg| \citep{quantreg}, our approach allows for flexible univariate forecast distributions, and a flexible impact of the regressors across different parts of the distribution. Our choice of realized measures as regressors for return quantiles follows \cite{ZikevsBarunik2015}. Our use of averages over different lags (one, five and $22$) follows \cite{Corsi2009} who proposes this construction as a simple approximation of long memory type persistence.

Having fitted the univariate forecast distributions, the approach described in step 4 is known as the Schaake shuffle. The latter was proposed by \cite{ClarkEtAl2004} as a pragmatic yet plausible device of pairing meteorological forecast distributions for various quantities. \cite{SchefzikEtAl2013} note its empirical copula interpretation. The key idea behind the Schaake shuffle is to draw univariate forecast distributions of the same size ($1000$) as the training sample, and then use the rank structure of the training sample to construct a multivariate forecast distribution. Here we use the rank structure of the standardized returns (Step 3); alternatively, the rank structure of the raw returns could be used. See \cite{GrotheEtAl2022} for a recent application of the Schaake shuffle to multivariate forecasting of energy prices and further discussion.

Table \ref{p_fin} shows our tests' results for the current case study. We focus on the energy score since a forecast density is not available in the current setup, rendering the use of the log score highly impractical. Auto-calibration of the univariate forecast distributions is not rejected at the $5 \%$ level, with a single exception (HSI, GBT test). Remarkably, auto-calibration of the ten-variate distribution for all assets is not rejected either, with both $p$-values exceeding $15 \%$ (last row of Table \ref{p_fin}). As a simple check of the tests' power, we also compute $p$-values for a na\"ive multivariate forecast distribution that pairs the univariate distributions in random order (i.e., assuming independence across return indexes). Reassuringly, both test versions clearly reject auto-calibration with a very small $p$-value below $0.1\%$.

\begin{figure}
\centering
\includegraphics{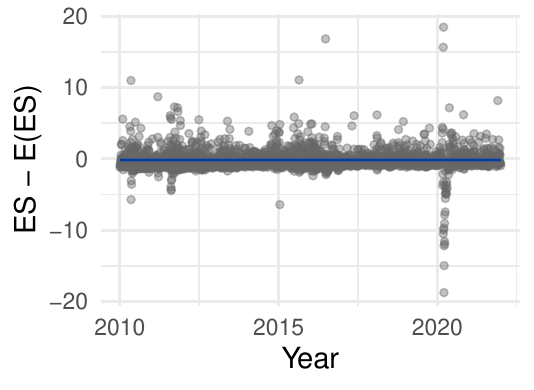}
\caption{Time series plot of difference between realized and expected energy scores ($\hat D_{\text{ES}, t}$) for the finance case study. Blue line shows a nonparametric time trend, based on local quadratic regression as implemented in the R function \textsf{loess}. \label{fig:Schaake}}
\end{figure}

Figure \ref{fig:Schaake} complements these results by plotting the entropy difference $\hat D_{\text{ES},t}$ over time. The flat nonparametric time trend (blue line) indicates that there is no meaningful time variation in either under- or overconfidence, indicating that the forecast model captures the dynamics of the  return distribution. Around the beginning of the Covid pandemic in March 2020, we observe a similar phenomenon as for the stochastic volatility model from Section 6: After a few extreme observations that come as a surprise to the model (large positive values for $\hat D_{\text{ES},t}$), we observe several large negative values for $\hat D_{\text{ES},t}$. These values can be explained by the model predicting large forecast uncertainty, in contrast to the realizing observations being moderate. However, this phenomenon is rather short-lived, and seems inevitable given the extreme (return and regressor) observations at the onset of the pandemic.

\begin{table}
\centering
\begin{tabular}{lrr}
	\toprule
	Index(es) & ES$_\text{D}$ & ES$_\text{GBT}$\\
	\midrule
DJI & 94.70 & 90.57\\
HSI & 16.92 & 0.79\\
BVSP & 95.65 & 62.61\\
FCHI & 95.66 & 25.21\\
FTSE & 45.62 & 57.34\\
\addlinespace
IBEX & 24.03 & 48.23\\
N225 & 33.20 & 33.73\\
SSMI & 18.75 & 47.48\\
GDAXI & 67.55 & 69.75\\
STOXX50E & 59.48 & 75.91\\
\addlinespace
\textbf{All (ten-variate)} & 16.55 & 59.08\\
\bottomrule
\end{tabular}
\caption{Test results ($p$-values in percent) for the finance case study. Test variants covered: ES$_\text{D}$ (energy score, entropy difference) and ES$_\text{GBT}$ (energy score, generalized Box transform). See text for details. \label{p_fin}}
\end{table}

\section{Conclusion} \label{sec:conclusion}

This paper proposes a framework for calibration testing of multivariate forecast distributions. This framework nests existing tests, and suggests improvements. We argue in favor of testing calibration based on proper scoring rules, and demonstrate that the resulting tests have good size and power properties in simulations. Furthermore, we show their usefulness in two challenging case studies from macroeconomics and finance. 

From a general perspective, our tests compare simulated model output to realized data and reject auto-calibration in case of large discrepancies. This broad idea of model checking is widely used across the sciences. For example, posterior predictive checks \citep[e.g.][Section 5]{Gabry2019} used in Bayesian statistics are based on this principle; see also \cite{CockayneEtAl2021} for a more theoretical perspective. In macroeconomics, structural economic models are often judged by how well they replicate certain features of empirical business cycle data. A common challenge across these (and many other) application areas is to decide just which `features of the data' are important and should thus be replicated by the forecast model. While this question may be answered by substantive concerns (such as a specific decision problem), forecast models are often produced for multiple purposes or for communication to a broader, diverse audience. This calls for a broadly applicable strategy to summarize data. We argue that proper scoring rules are well suited for this purpose, as evidenced by their wide use in forecast comparisons.

\clearpage

\singlespacing
\bibliography{calibration}

\begin{appendix}
	
\newpage

\section{Additional simulations}
\label{sec:appendix}
\subsection{Size}
\label{sec:addsim_size}

Here we explore the size of our proposed methods in two setups that are potentially more challenging than the one in Section \ref{sec:sim_baseline}: First, a setup involving time series dependence in forecasts and realizations. Second, a setup in which the true model follows a multivariate $t$ (rather than normal) distribution.

\subsubsection*{Time series dependence}

We simulate data from a $d$-dimensional vector autoregressive (VAR) model of order one:
$$Y_t = A~Y_{t-1} + \varepsilon_t,$$
where $Y_t$ and $\varepsilon_t$ are $d \times 1$ vectors of random variables with $d \in \{2, 4, 10\}$. All diagonal elements of the matrix $A$ are equal to $0.5$, and all off-diagonal elements are equal to zero. The error term vector $\varepsilon_t$ follows a multivariate normal distribution, with mean vector zero and covariance matrix $\gamma_t~\Sigma,$ with $\Sigma$ a $d\times d$ matrix and $\gamma_t$ being a scalar factor. $\Sigma$ has an equicorrelation structure with all diagonal elements equal to one and all off-diagonal elements equal to $0.5$. In the homoskedastic case, $\gamma_t = 1$ for all $t$; in the heteroskedastic case, $\gamma_t \in \{1, 1.25\}$ follows a first-order Markov Chain with two states. In each state, the probabilities of remaining and leaving are equal to $0.7$ and $0.3$ respectively. We consider forecast horizons of one and four periods, with the latter choice implying a three-period overlap in subsequent forecast errors. In order to account for possible autocorrelation under the null hypothesis, we use HAC approaches for implementing the $t$-test (used for the entropy tests) and the \cite{Knueppel2015} uniformity test (used for the GBT tests).\footnote{Our implementation of the uniformity test closely follows the test statistic called $\alpha_{1234}^0$ in \cite{Knueppel2015}. For the robust $t$-test, we use a HAC covariance estimator analogous to the one by \cite{Knueppel2015}.}

Table \ref{sim_size_var} presents simulation results for a sample of size $T = 200$. The rejection rates of all four test variants are between $0.04$ and $0.08$, and thus close to the nominal rate of $0.05$. These results indicate that time series dependence and the associated use of HAC approaches pose no severe challenges in the present setup.

\begin{table}[tbp]
	\centering
	\begin{tabular}{cccrrrr}
		\toprule
		$d$ & Horizon & Heterosk.? & ES$_\text{D}$ & ES$_\text{GBT}$ & LS$_\text{D}$ & LS$_\text{GBT}$\\
		\cmidrule{4-7}
		2 & 1 & No & 0.05 & 0.04 & 0.06 & 0.05\\
		4 & 1 & No & 0.06 & 0.05 & 0.05 & 0.05\\
		10 & 1 & No & 0.06 & 0.05 & 0.05 & 0.05\\ 	\addlinespace
		2 & 4 & No & 0.07 & 0.06 & 0.08 & 0.05\\
		4 & 4 & No & 0.07 & 0.05 & 0.07 & 0.05\\
		10 & 4 & No & 0.07 & 0.05 & 0.07 & 0.05\\ 	\addlinespace
		2 & 1 & Yes & 0.06 & 0.05 & 0.05 & 0.05\\
		4 & 1 & Yes & 0.06 & 0.05 & 0.06 & 0.05\\
		10 & 1 & Yes & 0.05 & 0.04 & 0.05 & 0.04\\ 	\addlinespace
		2 & 4 & Yes & 0.07 & 0.05 & 0.07 & 0.05\\
		4 & 4 & Yes & 0.06 & 0.05 & 0.07 & 0.05\\
		10 & 4 & Yes & 0.08 & 0.05 & 0.07 & 0.05\\
		\bottomrule
	\end{tabular}
	\caption{Size under time series dependence with $T=200$. Nominal size equals $0.05$. Rejection rates are based on $5000$ simulations. \label{sim_size_var}}
\end{table}

\subsubsection*{$t$ distribution}

We additionally consider a static setup where $Y_t$ follows a $d$-variate $t$ distribution with eight degrees of freedom. We consider $d \in \{2,4,10\}$ and a sample size of $T = 200$. Table \ref{sim_size_t} shows that the rejection rates are between $0.05$ and $0.06$ in all cases, again indicating good size properties.

\begin{table}[tbp]
	\centering
	\begin{tabular}{rrrrr}
		\toprule
		$d$ &  ES$_\text{D}$ & ES$_\text{GBT}$ & LS$_\text{D}$ & LS$_\text{GBT}$\\
		\cmidrule{2-5}
2 & 0.06 & 0.05 & 0.06 & 0.05\\
4 & 0.06 & 0.06 & 0.06 & 0.05\\
10 & 0.05 & 0.05 & 0.05 & 0.05\\
		\bottomrule
	\end{tabular}
	\caption{Size in a $t$-distributed setup ($10$ degrees of freedom) with $T=200$. Nominal size equals $0.05$. Rejection rates are based on $5000$ simulations. \label{sim_size_t}}
\end{table}

\subsection{Additional simulations: Power}

Table \ref{tab:power} presents further simulation results on the power of the tests. While the true data is simulated from the same VAR model as in Section \ref{sec:addsim_size}, we consider various types of deviations of the forecast model from the true model. The left half of Table \ref{tab:power} lists the specific setups considered. As expected, all tests have higher power for the larger sample size $T = 200$, as compared to $T = 50$. Furthermore, we generally observe a slight decrease in power when moving to a higher forecast horizon ($h = 4$ instead of $h = 1$), which seems plausible given the overlap and increased persistence in the relevant calibration statistics. As in the baseline simulations of Section \ref{sec:sim_baseline}, the tests based on the log score have somewhat higher power than the ones using the energy score. The entropy variant of the test tends to have higher power than the GBT variant in the dynamic settings considered here, which is likely due to fact that all settings imply wrong variance forecasts, corresponding to setting $H_1$ in Table \ref{tab:DovMan}.

\begin{table}[tbp]
	\centering
	\begin{tabular}{rrllrrrlrrrr}
		\toprule
		& & \multicolumn{2}{c}{Heterosk.?} & \multicolumn{2}{c}{$\Sigma$} & \multicolumn{2}{c}{$A$} & \multicolumn{4}{c}{Rejection Rates} \\ 	\cmidrule{3-8}
		
		$T$ & $h$ & True & Fcst & True & Fcst & True & Fcst & ES$_\text{D}$ & ES$_\text{GBT}$ & LS$_\text{D}$ & LS$_\text{GBT}$\\
		\cmidrule{9-12}
50 & 1 & Yes & Yes & 1 & 1.0 & 0.5 & 0.8 & 0.13 & 0.05 & 0.17 & 0.06\\
200 & 1 & Yes & Yes & 1 & 1.0 & 0.5 & 0.8 & 0.39 & 0.23 & 0.55 & 0.33\\
50 & 4 & Yes & Yes & 1 & 1.0 & 0.5 & 0.8 & 0.14 & 0.05 & 0.17 & 0.05\\
200 & 4 & Yes & Yes & 1 & 1.0 & 0.5 & 0.8 & 0.33 & 0.20 & 0.45 & 0.27\\ \addlinespace
50 & 1 & Yes & Yes & 1 & 1.2 & 0.5 & 0.5 & 0.46 & 0.14 & 0.82 & 0.25\\
200 & 1 & Yes & Yes & 1 & 1.2 & 0.5 & 0.5 & 0.91 & 0.82 & 1.00 & 0.99\\
50 & 4 & Yes & Yes & 1 & 1.2 & 0.5 & 0.5 & 0.41 & 0.09 & 0.68 & 0.14\\
200 & 4 & Yes & Yes & 1 & 1.2 & 0.5 & 0.5 & 0.76 & 0.61 & 0.99 & 0.88\\ \addlinespace
50 & 1 & Yes & No & 1 & 1.0 & 0.5 & 0.5 & 0.12 & 0.05 & 0.17 & 0.06\\
200 & 1 & Yes & No & 1 & 1.0 & 0.5 & 0.5 & 0.42 & 0.25 & 0.57 & 0.35\\
50 & 4 & Yes & No & 1 & 1.0 & 0.5 & 0.5 & 0.11 & 0.04 & 0.14 & 0.04\\
200 & 4 & Yes & No & 1 & 1.0 & 0.5 & 0.5 & 0.29 & 0.16 & 0.42 & 0.25\\
		\bottomrule
	\end{tabular}
	\caption{Power under time series dependence. Here auto-calibration is violated, and rejection rates represent power. Rows 1-4 correspond to misspecified persistence (diagonal elements of $A$ equal to $0.8$ instead of $0.5$). Rows 5-8 correspond to misspecified residual variance (diagonal elements of $\Sigma$ equal to $1.2$ instead of $1.0$). Rows 9-12 correspond to neglected heteroskedasticity. Rejection rates are based on $5000$ simulations.} \label{tab:power}
\end{table}

\end{appendix}

\end{document}

%% file: dmcompare6.tex
\begin{tabular}
{rcccccccccccccc}
\hline\hline
& &  $Z_t^{2\dagger}$ & AvR  &  ES\textsubscript{D}  &  ES\textsubscript{GBT}  &  LS\textsubscript{D}  &  LS\textsubscript{GBT}   &  & $Z_t^{2\dagger}$   &  AvR &  ES\textsubscript{D}  &  ES\textsubscript{GBT}  &  LS\textsubscript{D}  &  LS\textsubscript{GBT}   \\

$d$    & &   \multicolumn{6}{c}{$T=50$}   &   & \multicolumn{6}{c}{$T=200$} \\ \cline{3-8}  \cline{10-15}

\multicolumn{15}{c}{}        \\

          &       & \multicolumn{13}{c}{Size} \\

    2     &       & 0.05  & 0.06  & 0.07  & 0.05  & 0.07  & 0.05  &       & 0.06  & 0.05  & 0.05  & 0.05  & 0.05  & 0.05 \\
    4     &       & 0.05  & 0.05  & 0.06  & 0.05  & 0.05  & 0.05  &       & 0.06  & 0.06  & 0.05  & 0.05  & 0.05  & 0.05 \\
    6     &       & 0.05  & 0.05  & 0.06  & 0.05  & 0.06  & 0.05  &       & 0.05  & 0.05  & 0.05  & 0.04  & 0.04  & 0.05 \\
    10    &       & 0.06  & 0.05  & 0.06  & 0.05  & 0.05  & 0.06  &       & 0.06  & 0.05  & 0.06  & 0.05  & 0.06  & 0.05 \\
    20    &       & 0.05  & 0.05  & 0.06  & 0.06  & 0.05  & 0.05  &       & 0.05  & 0.05  & 0.06  & 0.05  & 0.05  & 0.05 \\
    50    &       & 0.05  & 0.05  & 0.06  & 0.06  & 0.05  & 0.05  &       & 0.05  & 0.05  & 0.05  & 0.05  & 0.05  & 0.05 \\
          &       & \multicolumn{13}{c}{Power given data from $H_1$ (larger variance)} \\
    2     &       & 0.12  & 0.05  & 0.16  & 0.11  & 0.17  & 0.13  &       & 0.44  & 0.14  & 0.65  & 0.45  & 0.71  & 0.51 \\
    4     &       & 0.21  & 0.05  & 0.25  & 0.17  & 0.37  & 0.24  &       & 0.79  & 0.12  & 0.85  & 0.69  & 0.95  & 0.86 \\
    6     &       & 0.31  & 0.05  & 0.31  & 0.22  & 0.54  & 0.36  &       & 0.94  & 0.11  & 0.92  & 0.81  & 0.99  & 0.97 \\
    10    &       & 0.55  & 0.05  & 0.37  & 0.30  & 0.80  & 0.59  &       & 1.00  & 0.11  & 0.97  & 0.93  & 1.00  & 1.00 \\
    20    &       & 0.90  & 0.05  & 0.44  & 0.45  & 0.98  & 0.91  &       & 1.00  & 0.10  & 0.99  & 0.99  & 1.00  & 1.00 \\
    50    &       & 1.00  & 0.05  & 0.51  & 0.76  & 1.00  & 1.00  &       & 1.00  & 0.10  & 1.00  & 1.00  & 1.00  & 1.00 \\
          &       & \multicolumn{13}{c}{Power given data from $H_2$ (smaller correlation)} \\
    2     &       & 0.05  & 0.08  & 0.06  & 0.07  & 0.05  & 0.05  &       & 0.17  & 0.11  & 0.04  & 0.07  & 0.11  & 0.08 \\
    4     &       & 0.16  & 0.12  & 0.05  & 0.10  & 0.15  & 0.10  &       & 0.63  & 0.26  & 0.05  & 0.17  & 0.57  & 0.39 \\
    6     &       & 0.25  & 0.16  & 0.05  & 0.13  & 0.28  & 0.17  &       & 0.89  & 0.37  & 0.06  & 0.31  & 0.88  & 0.72 \\
    10    &       & 0.50  & 0.19  & 0.05  & 0.20  & 0.60  & 0.37  &       & 0.99  & 0.47  & 0.08  & 0.58  & 1.00  & 0.97 \\
    20    &       & 0.86  & 0.20  & 0.05  & 0.36  & 0.94  & 0.81  &       & 1.00  & 0.56  & 0.09  & 0.91  & 1.00  & 1.00 \\
    50    &       & 1.00  & 0.23  & 0.05  & 0.67  & 1.00  & 1.00  &       & 1.00  & 0.61  & 0.11  & 1.00  & 1.00  & 1.00 \\
          &       & \multicolumn{13}{c}{Power given data from $H_3$ ($t(8)$-distribution)} \\
    2     &       & 0.10  & 0.09  & 0.11  & 0.10  & 0.10  & 0.11  &       & 0.39  & 0.14  & 0.11  & 0.39  & 0.07  & 0.47 \\
    4     &       & 0.22  & 0.09  & 0.11  & 0.19  & 0.08  & 0.28  &       & 0.88  & 0.14  & 0.12  & 0.80  & 0.06  & 0.94 \\
    6     &       & 0.43  & 0.08  & 0.11  & 0.30  & 0.08  & 0.50  &       & 0.99  & 0.14  & 0.15  & 0.95  & 0.07  & 1.00 \\
    10    &       & 0.79  & 0.09  & 0.11  & 0.52  & 0.08  & 0.84  &       & 1.00  & 0.14  & 0.14  & 1.00  & 0.06  & 1.00 \\
    20    &       & 0.99  & 0.09  & 0.13  & 0.81  & 0.09  & 1.00  &       & 1.00  & 0.13  & 0.15  & 1.00  & 0.07  & 1.00 \\
    50    &       & 1.00  & 0.08  & 0.13  & 0.97  & 0.08  & 1.00  &       & 1.00  & 0.13  & 0.15  & 1.00  & 0.06  & 1.00 \\
          &       & \multicolumn{13}{c}{Power given data from $H_4$ (CCC-GARCH(1,1))} \\
    2     &       & 0.26  & 0.20  & 0.43  & 0.28  & 0.44  & 0.31  &       & 0.39  & 0.27  & 0.48  & 0.43  & 0.48  & 0.46 \\
    4     &       & 0.28  & 0.17  & 0.44  & 0.29  & 0.45  & 0.33  &       & 0.48  & 0.21  & 0.50  & 0.45  & 0.52  & 0.56 \\
    6     &       & 0.32  & 0.15  & 0.45  & 0.30  & 0.47  & 0.37  &       & 0.56  & 0.20  & 0.48  & 0.49  & 0.52  & 0.65 \\
    10    &       & 0.41  & 0.15  & 0.45  & 0.36  & 0.52  & 0.46  &       & 0.73  & 0.18  & 0.50  & 0.54  & 0.56  & 0.79 \\
    20    &       & 0.54  & 0.14  & 0.46  & 0.40  & 0.58  & 0.57  &       & 0.90  & 0.18  & 0.50  & 0.62  & 0.60  & 0.92 \\
    50    &       & 0.76  & 0.13  & 0.46  & 0.47  & 0.66  & 0.77  &       & 1.00  & 0.18  & 0.53  & 0.80  & 0.71  & 1.00 \\

\hline\hline
\multicolumn{15}{ p{15.5cm}}{\scriptsize{Note: $Z_t^{2\dagger}$ denotes the test proposed by \cite{DovernManner2020}, AvR the average rank test based on \cite{ThorarinsdottirEtAl2016}, ES the tests based on the energy score and LS the tests based on the log score. The subscript D refers to the entropy variant of the tests, the subscript GBT to the generalized Box transform variant. $d$ denotes the number of variables and $T$ the number of periods. The rejection rates are obtained using 5000 simulations, a standard $t$-test in the cases of ES\textsubscript{D} and LS\textsubscript{D}, and the raw-moments test by \cite{Knueppel2015} with zero lags in all other cases. $J=J_0=J_1=5000$ for all simulation-based tests.}} \\
\end{tabular} 